\documentclass[12pt]{article}
\usepackage{amsmath}
\usepackage{graphicx}
\usepackage{amssymb}
\usepackage[all]{xy}
\usepackage{pdflscape}
\renewcommand{\theequation}{\arabic{section}.\arabic{equation}}

\begin{document}



\def\a{\alpha}
\def\b{\beta}
\def\d{\delta}
\def\e{\epsilon}
\def\g{\gamma}
\def\h{\mathfrak{h}}
\def\k{\kappa}
\def\l{\lambda}
\def\o{\omega}
\def\p{\wp}
\def\r{\rho}
\def\t{\tau}
\def\s{\sigma}
\def\z{\zeta}
\def\x{\xi}
\def\V={{{\bf\rm{V}}}}
 \def\A{{\cal{A}}}
 \def\B{{\cal{B}}}
 \def\C{{\cal{C}}}
 \def\D{{\cal{D}}}
\def\K{{\cal{K}}}
\def\O{\Omega}
\def\R{\bar{R}}
\def\T{{\cal{T}}}
\def\L{\Lambda}
\def\f{E_{\tau,\eta}(sl_2)}
\def\E{E_{\tau,\eta}(sl_n)}
\def\Zb{\mathbb{Z}}
\def\Cb{\mathbb{C}}

\def\R{\overline{R}}

\def\beq{\begin{equation}}
\def\eeq{\end{equation}}
\def\bea{\begin{eqnarray}}
\def\eea{\end{eqnarray}}
\def\ba{\begin{array}}
\def\ea{\end{array}}
\def\no{\nonumber}
\def\le{\langle}
\def\re{\rangle}
\def\lt{\left}
\def\rt{\right}

\newtheorem{Theorem}{Theorem}
\newtheorem{Definition}{Definition}
\newtheorem{Proposition}{Proposition}
\newtheorem{Lemma}{Lemma}
\newtheorem{Corollary}{Corollary}
\newcommand{\proof}[1]{{\bf Proof. }
        #1\begin{flushright}$\Box$\end{flushright}}

\baselineskip=20pt

\newfont{\elevenmib}{cmmib10 scaled\magstep1}
\newcommand{\preprint}{
   \begin{flushleft}
   \end{flushleft}\vspace{-1.3cm}
   \begin{flushright}\normalsize
   \end{flushright}}
\newcommand{\Title}[1]{{\baselineskip=26pt
   \begin{center} \Large \bf #1 \\ \ \\ \end{center}}}
\newcommand{\Author}{\begin{center}
   \large \bf
Kun Hao${}^{a,b}$,~Fakai Wen${}^{a,c}$,~Junpeng Cao${}^{c,d}$,~Guang-Liang Li${}^{e}$,~Wen-Li Yang${}^{a,b,f}\footnote{Corresponding author:
wlyang@nwu.edu.cn}$,
 ~ Kangjie Shi${}^{a,b}$ and~Yupeng Wang${}^{c,d}\footnote{Corresponding author: yupeng@iphy.ac.cn}$
 \end{center}}
\newcommand{\Address}{\begin{center}

     ${}^a$Institute of Modern Physics, Northwest University,
     Xian 710069, China\\
     ${}^b$Shaanxi Key Laboratory for Theoretical Physics Frontiers,  Xian 710069, China\\
     ${}^c$Beijing National Laboratory for Condensed Matter
           Physics, Institute of Physics, Chinese Academy of Sciences, Beijing
           100190, China\\
     ${}^d$Collaborative Innovation Center of Quantum Matter, Beijing,
     China\\
     ${}^e$Department of Applied Physics, Xian Jiaotong University, Xian 710049, China\\
     ${}^f$Beijing Center for Mathematics and Information Interdisciplinary Sciences, Beijing, 100048,  China

   \end{center}}

\preprint \thispagestyle{empty}
\bigskip\bigskip\bigskip

\Title{On the solutions of the $Z_n$-Belavin model with arbitrary number of sites} \Author

\Address \vspace{1cm}

\begin{abstract}
The periodic $Z_n$-Belavin model on a lattice with an arbitrary number of sites $N$  is studied via the off-diagonal Bethe
Ansatz  method (ODBA). The eigenvalues of the corresponding transfer matrix are given in terms of an unified inhomogeneous $T-Q$ relation. In the special
case of $N=nl$ with $l$ being also a positive integer, the resulting $T-Q$ relation recovers the
homogeneous one  previously  obtained via algebraic Bethe Ansatz.

\vspace{1truecm} \noindent {\it PACS:} 75.10.Pq, 03.65.Vf, 71.10.Pm


\end{abstract}

\newpage



\section{Introduction}
\label{intro} \setcounter{equation}{0}

Our understanding to phase transitions and critical phenomena has been greatly
enhanced by the study on lattice integrable models \cite{Bax82}. Such exact results
provide valuable insights into the key theoretical development of universality classes in areas
ranging from modern condensed physics \cite{Duk04,Gua13} to string and super-symmetric Yang-Mills theories
\cite{Mad99,Dol03,Bei12}. Among solvable models \cite{Bax82,Kor93,McC10}, elliptic ones stand out as a
particularly important class due to the fact that most others can be reduced to from them by taking
trigonometric or rational limits. The $Z_n$-Belavin model \cite{Bel81} is a typical elliptic quantum integrable model,
with the celebrated XYZ spin chain as the special case of $n=2$.

The first exact solution of the $Z_2$-model with periodic boundary
condition was given by Baxter \cite{Bax71}, where the fundamental equation (the
Yang-Baxter equation \cite{Bax82,Yan67}) was emphasized and the $T-Q$ method was proposed.
Takhtadzhan and Faddeev \cite{Tak79} resolved the model with the algebraic Bethe Ansatz method
\cite{Kor93, Skl78}. By employing the intertwiners vectors \cite{Jim87} which constitute the
face-vertex correspondence between the  $Z_n$-Belavin model and  the associated face model,
Hou et al \cite{Hou89}  generalized Takhatadzhan and Faddeev's approach to the $Z_n$-Belavin
model with a generic $n$.
In their approach, local gauge transformation played a central role
to obtain local vacuum states (reference states) with
which the algebraic Bethe Ansatz analysis can be performed. However, such reference states are so far only available
for some very particular number of lattice sites,  namely,  $N=nl$  with $l$ being a positive integer,  but not for the other $N$.
This leads to the fact that the conventional Bethe Ansatz methods have been quite hard to apply to the latter case for many years.
In fact, the lack of a reference state is a common feature of the integrable models without $U(1)$ symmetry and
had been a very important and difficult issue in the field of quantum integrable models.

Recently, a systematic method, i.e.,  the off-diagonal Bethe Ansatz (ODBA) \cite{Cao13,Wan15} was proposed to solve the eigenvalue problem of integrable models without U(1)-symmetry. The
closed XYZ spin chain (or the $Z_2$-model) with arbitrary number of sites \cite{Cao14} and several  other long-standing models
\cite{Cao13, Cao14JHEP143,Hao14,Cao15} have since been solved. In this paper, we adopt ODBA to solve the eigenvalue problem of the periodic $Z_n$-Belavin model with a generic positive
integer $n\geq 2$ and an arbitrary lattice number $N$.

The paper is organized as follows. Section 2 serves as an introduction of our notations and some basic ingredients.
The commuting transfer matrix associated with the periodic $Z_n$-Belavin model is  constructed to show the integrability of the model.
In section 3, based on some intrinsic properties of the $Z_n$-Belavin's $R$-matrix, we construct the fused transfer matrices
by anti-symmetric fusion procedure  and   derive some operator identities
and the quasi-periodicities of these matrices. Taking the $Z_3$ model
as a concrete example,  we express the eigenvalues of the transfer matrix in terms of a nested inhomogeneous $T-Q$ relation  and
the associated Bethe Ansatz equations (BAEs) in Section 4. Generalization
to $Z_n$ case  is presented in Section 5. We summarize our results and give some discussions in Section 6.
A slightly detailed description about the $Z_4$ case, which might be crucial to
understand the procedure for $n\geq4$, is given in Appendix A.
In addition, we discuss the ODBA solution of $Z_n$-Belavin model with twisted boundary condition in Appendix B.


\section{ $Z_n$-Belavin model with periodic boundary condition}
\label{Zn} \setcounter{equation}{0}

Let us fix a positive integer $n\geq 2$, a complex number $\tau$
such that $Im(\tau)>0$ and a generic complex number $w$. For convenience, let us introduce
the elliptic functions
\begin{eqnarray}
\theta\lt[
 \begin{array}{c}
 a\\b
 \end{array}\rt](u,\tau)&=&\sum_{m=-\infty}^{\infty}
 exp\lt\{\sqrt{-1}\pi\lt[(m+a)^2\tau+2(m+a)(u+b)\rt]\rt\},\\[4pt]
 \theta^{(j)}(u)&=&\theta\lt[\begin{array}{c}\frac{1}{2}-\frac{j}{n}\\
 [2pt]\frac{1}{2}
 \end{array}\rt](u,n\tau),\qquad
 \s(u)=\theta\lt[\begin{array}{c}\frac{1}{2}\\[2pt]\frac{1}{2}
 \end{array}\rt](u,\tau),\label{Function}\\[4pt]
 \zeta(u)&=&\frac{\partial}{\partial u}\lt\{\ln
 \s(u)\rt\}.\no
\end{eqnarray}
Among them the
$\s$-function\footnote{Our $\s$-function is the
$\vartheta$-function $\vartheta_1(u)$ \cite{Whi50}. It has the
following relation with the {\it Weierstrassian\/} $\s$-function
denoted by
$\s_w(u)$: $\s_w(u)\propto e^{\eta_1u^2}\s(u)$,
$\eta_1=\pi^2(\frac{1}{6}-4\sum_{n=1}^{\infty}\frac{nq^{2n}}{1-q^{2n}})
$ and $q=e^{\sqrt{-1}\tau}$. }
 satisfies the following
identity:
\begin{eqnarray}
 &&\s(u+x)\s(u-x)\s(v+y)\s(v-y)-\s(u+y)\s(u-y)\s(v+x)\s(v-x)\no\\[4pt]
 &&\qquad\qquad \qquad =\s(u+v)\s(u-v)\s(x+y)\s(x-y).\no
\end{eqnarray}

Let ${\rm\bf V}$ denote an $n$-dimensional linear space with an orthonormal basis $\{|i\rangle|i=1,\cdots,n\}$, and  $g,\,h$ be two
$n\times n$ matrices with the elements
\begin{eqnarray}
 &&h_{ij}=\d_{i+1\,j},~~g_{ij}={\o_n}^i\d_{i\,j},~~{\rm
 with}\quad \o_n=e^{\frac{2\pi\sqrt{-1}}{n}},~~~i,j\in \Zb_n,\no
\end{eqnarray}
namely,
\bea
g=\lt(\begin{array}{ccccc}1&&&&\\
&\o_n&&&\\&&{\o_n}^2&&\\
&&&\ddots&\\
&&&&{\o_n}^{n-1}\end{array}\rt),\quad
h=\lt(\begin{array}{ccccc}&1&&&\\
&&\ddots&&\\
&&&\ddots&\\
&&&&1\\
1&&&& \end{array}\rt).
\eea
It is easy to verify that the matrices satisfy the relation
\bea
gh={\o_n}^{-1}\,hg.\label{gh-relation}
\eea
Associated with  any ${\bf \a}=(\a_1,\a_2)$, $\a_1,\,\a_2\in \Zb_n$, one can introduce an
$n\times n$ matrix $I_{{\bf \a}}$  defined  by
\begin{eqnarray}
 &&I_{{\bf \a}}=I_{(\a_1,\a_2)}=g^{\a_2}h^{\a_1},\label{I-matrix}
\end{eqnarray}
and an elliptic function $\s_{{\bf \a}}(u)$ given by
\begin{eqnarray}
 &&\s_{{\bf \a}}(u)=\theta\lt[\begin{array}{c}\frac{1}{2}+\frac{\a_1}{n}\\[4pt]
 \frac{1}{2}+\frac{\a_2}{n}
 \end{array}\rt](u,\tau),~~{\rm and\/}~~\s_{(0,0)}(u)=\s(u).\no
\end{eqnarray}
The $\Zb_n$-Belavin R-matrix  $R(u)\in {\rm End}({\rm\bf V}\otimes
{\rm\bf V})$  is given by \cite{Bel81,Ric86,Jim87}
\begin{eqnarray}
 R(u)=\sum_{{\bf \a}\in\Zb_n^2}
 \frac{\s_{{\bf \a}}(u+\frac{w}{n})}
 {n\s_{{\bf \a}}(\frac{w}{n})}I_{{\bf \a}}\otimes
 I_{{\bf \a}}^{-1},\label{Belavin-R}
\end{eqnarray}
which satisfies the
quantum Yang-Baxter equation (QYBE)
\begin{eqnarray}
 R_{12}(u_1-u_2)R_{13}(u_1-u_3)R_{23}(u_2-u_3)=
 R_{23}(u_2-u_3)R_{13}(u_1-u_3)R_{12}(u_1-u_2), \label{QYB}
\end{eqnarray}
and the properties \cite{Ric86},
\begin{eqnarray}
 &&\hspace{-1.45cm}\mbox{
 Initial condition}:\hspace{42.5mm}R_{12}(0)=  P_{1,2},\label{Initial}\\[6pt]
 &&\hspace{-1.5cm}\mbox{
 Unitarity}:\hspace{42.5mm}R_{12}(u)R_{21}(-u)= \frac{\s(w+u)\,\s(w-u)}{\s(w)\,\s(w)}\times{\rm id},\label{Unitarity}\\[6pt]
 &&\hspace{-1.5cm}\mbox{
 Crossing-unitarity}:\quad R^{t_2}_{21}(-u-nw)\,R_{12}^{t_2}(u)
 = \frac{\s(u)\s(-u-nw)}{\s(w)\s(w)}\,\times\mbox{id},
 \label{crosing-unitarity}\\[6pt]
 &&\hspace{-1.5cm}\mbox{
 $Z_n$-symmetry}:\quad g_1\,g_2\,R_{12}(u)g_1^{-1}\,g_2^{-1}=R_{12}(u),\quad
  h_1\,h_2\,R_{12}(u)h_1^{-1}\,h_2^{-1}=R_{12}(u),
 \label{Zn-symmetry}\\[6pt]
 &&\hspace{-1.4cm}\mbox{Fusion conditions}:\hspace{22.5mm}\, R_{12}(-w)= P^{(-)}_{1,2}\, S^{(-)}_{12}, \quad
 R_{12}(w)=S^{(+)}_{12}\, P^{(+)}_{1,2}.\label{Fusion}
\end{eqnarray}
Here $R_{21}(u)=P_{1,2}R_{12}(u)P_{1,2}$ with $P_{1,2}$
being the usual permutation operator, $P^{(\mp)}_{1,2}=\frac{1}{2}\{1\mp P_{1,2}\}$ is anti-symmetric
(symmetric) project operator in the tensor product space  ${\rm\bf
V} \otimes {\rm\bf V} $, $S^{(\pm)}_{12}$ are some non-degenerate matrices  $\in {\rm End}({\rm\bf V}\otimes
{\rm\bf V})$ \cite{Ric86,Jim87} and $t_i$ denotes the
transposition in the $i$-th space. Here and below we adopt the
standard notation: for any matrix $A\in {\rm End}({\rm\bf V})$, $A_j$
is an embedding operator in the tensor space ${\rm\bf V}\otimes
{\rm\bf V}\otimes\cdots$, which acts as $A$ on the $j$-th space and as
an identity on the other factor spaces; $R_{ij}(u)$ is an
embedding operator of R-matrix in the tensor space, which acts as
an identity on the factor spaces except for the $i$-th and $j$-th
ones.

As usual, the corresponding ``row-to-row" monodromy matrix $T(u)$
\cite{Kor93}, an $n\times n$ matrix with operator-valued elements  acting  on $({\rm\bf V})^{\otimes N}$ reads
\begin{eqnarray}
T_0(u)=R_{0N}(u-\theta_N)R_{0N-1}(u-\theta_{N-1})\cdots
R_{01}(u-\theta_1).\label{T-matrix}
\end{eqnarray}
Here $\{\theta_i|i=1,\cdots, N\}$
are arbitrary free complex parameters which are usually called
the inhomogeneous parameters. With the help of the QYBE (\ref{QYB}),
one can show that $T(u)$ satisfies the Yang-Baxter algebra relation
\begin{eqnarray}
R_{12}(u-v)T_1(u)T_2(v)=T_2(v)T_1(u)R_{12}(u-v).\label{Relation1}
\end{eqnarray}
Let us introduce the transfer matrix $t(u)$
\bea
t(u)=tr_0\lt(T_0(u)\rt)=tr\lt(T(u)\rt).\label{transfer}
\eea
The $Z_n$-Belavin model \cite{Bel81} with periodic boundary condition is a quantum spin chain described by
the Hamiltonian
\bea
H=\frac{\partial}{\partial u}\lt.\lt\{\ln t(u)\rt\}\rt|_{u=0,\{\theta_i\}=0}-{\rm Constant}=\sum_{i=1}^NH_{i,i+1},\label{Ham}
\eea
where the local Hamiltonian  $H_{i,i+1}$ is
\bea
H_{i,i+1}= \frac{\partial}{\partial u}\lt.\lt\{P_{i,i+1} R_{i,i+1}(u)\rt\}\rt|_{u=0},
\eea
with the periodic boundary condition, namely,
\bea
H_{N,N+1}=H_{N,1}.\label{Periodic-boundary}
\eea
The commutativity of the transfer matrices
\begin{eqnarray}
 [t(u),t(v)]=0,\no
\end{eqnarray}
follows as a consequence of (\ref{Relation1}). This ensures the integrability of
the inhomogeneous $Z_n$-Belavin model with periodic boundary.


\section{Relations of the eigenvalues}
\setcounter{equation}{0}
Following the method developed in \cite{Cao14JHEP143} (see also Chapter 7 of \cite{Wan15}), we apply the fusion techniques \cite{Kar79,Kulish81,Kulish82,Kirillov86} to study the $Z_n$-Belavin model. Besides the fundamental transfer
matrix $t(u)$ some other fused transfer matrices $\{t_j(u)|j=1,\cdots,n\}$ (see below (\ref{transfer-fused})), which commute with each other and include
the original one as  $t_1(u)=t(u)$, are constructed through an anti-symmetric fusion procedure with the help of the fusion condition (\ref{Fusion}) of the $R$-matrix.

\subsection{Operator product identities}
The quasi-periodicity of the $\s$-function
\bea
\s(u+1)=-\s(u),\quad \s(u+\tau)=-e^{-2i\pi(u+\frac{\tau}{2})}\,\s(u),\label{quasi-periodic}
\eea
indicates that the $R$-matrix $R(u)$ given by (\ref{Belavin-R}) possesses the quasi-periodic properties
\bea
&&R_{12}(u+1)=-g_1^{-1}\,R_{12}(u)\,g_1=-g_2\,R_{12}(u)\,g_2^{-1},\label{Quasi-R-1}\\[4pt]
&&R_{12}(u+\tau)=-e^{-2i\pi(u+\frac{w}{n}+\frac{\tau}{2})}\,h_1^{-1}\,R_{12}(u)\,h_1
=-e^{-2i\pi(u+\frac{w}{n}+\frac{\tau}{2})}\,h_2\,R_{12}(u)\,h_2^{-1},\label{Quasi-R-2}
\eea
which lead to the quasi-periodicity of the transfer matrix $t(u)$ given by (\ref{transfer})
\bea
&&t(u+1)=(-1)^N \,t(u),\label{Quasi-transfer-1}\\[4pt]
&&t(u+\tau)=(-1)^N\,e^{-2i\pi\{N(u+\frac{w}{n}+\frac{\tau}{2})-\sum_{l=1}^N\theta_l\}}\,t(u).\label{Quasi-transfer-2}
\eea

Let us introduce the usual (or non-deformed)  anti-symmetric projectors $\{P^{(-)}_{1,\cdots,m}|m = 2, . . . , n\}$ in
a tensor space of V defined by the induction relations
\bea
P^{(-)}_{1,\cdots,m+1}=\frac{1}{m+1}(1-\sum_{j=2}^{m+1}P_{1,j})P^{(-)}_{2,\cdots,m+1},\quad m=2,\cdots,n-1.\no
\eea
Iterating the above relation yields alternative definition of the projectors
\bea
P^{(-)}_{1,\cdots,m} =\frac{1}{m!}\sum_{\kappa\in S_m}\,
(-1)^{sign(\kappa)}P_{\kappa} , m = 2,\cdots, n,\label{Projectors}
\eea
where $S_m$ is the permutation group of $m$ indices, $P_{\kappa}$ is a permutation in the group,
and $sign(\kappa)$ is $0$ for an even permutation $\kappa$ and $1$ for an odd permutation. With the above anti-symmetric
projectors, we can construct the fused monodromy matrices
\bea
T_{\langle 1,\cdots,m\rangle}(u)= P^{(-)}_{1,\cdots,m}\, T_1(u)\cdots T(u-(m-1)w)\,P^{(-)}_{1,\cdots,m},\quad m=2,\cdots,n.
\label{Fusion-Mon}
\eea
The corresponding fused transfer matrices $\{t_j(u)|j=1,\cdots,n\}$ (including
the original one as  $t_1(u)=t(u)$) are then given by
\bea
t_m(u)=tr_{1,\cdots,m}\lt\{T_{\langle 1,\cdots,m\rangle}(u)\rt\}, \quad m=2,\cdots,n. \label{transfer-fused}
\eea
The last fused transfer matrix $t_n(u)$ is the so-called quantum determinant \cite{Kul79}
which plays the role of the generating functional of the centers of the associated quantum
algebras \cite{Cha94}. For generic values of $\{\theta_j\}$, $t_n(u)$ is proportional to the identity operator, namely,
\bea
t_n(u)={\rm Det}_q\lt(T(u)\rt)\times {\rm id},\quad {\rm Det}_q\lt(T(u)\rt)=\prod_{l=1}^N\frac{\s(u-\theta_l+w)}{\s(w)}
\,\prod_{k=1}^{n-1}\frac{\s(u-\theta_l-kw)}{\s(w)}.\label{quant-deter}
\eea

The QYBE (\ref{QYB}),  the fusion condition (\ref{Fusion}) and the relations (\ref{Relation1}) imply that
these fused matrices commute with each other,
\bea
[t_i(u),\,t_j(v)]=0,\quad i,j=1,\cdots,n.\label{commutive}
\eea
Using the method (see  Chapter 7.2 of \cite{Wan15}), we can show the  relations
\bea
T_1(\theta_j)\,T_{\langle 2,\cdots,m\rangle}(\theta_j-w)=P^{(-)}_{1,\cdots,m}T_1(\theta_j)\,T_{\langle 2,\cdots,m\rangle}(\theta_j-w),
\,\, m=2,\cdots,n;\,\,j=1,\cdots,N,\no
\eea
which immediately lead to the following recursive relations
\bea
t(\theta_j)\,t_m(\theta_j-w)=t_{m+1}(\theta_j),\quad m=1,\cdots,n-1;\quad j=1,\cdots,N.\label{Oper-1}
\eea
Moreover, the fusion condition (\ref{Fusion}) and the fact that $P^{(-)}_{12}\,P^{(+)}_{12}=
0=P^{(+)}_{12}\,P^{(-)}_{12}$ enable us to derive some zeros for the fused matrices,
\bea
t_m(\theta_j+kw)=0,\quad \quad j=1,\cdots,N;\,\, k=1,\cdots,m-1;\,\,m=2,\cdots,n.\label{Zeros-Op}
\eea
Similarly as deriving the relations (\ref{Quasi-transfer-1})-(\ref{Quasi-transfer-2}), we have the fused transfer
matrices enjoy the following periodicity,
\bea
\hspace{-1.2truecm}&&t_m(u+1)=(-1)^{mN} \,t_m(u),\quad m=1,\cdots,n,\label{Quasi-transfer-3}\\[4pt]
\hspace{-1.2truecm}&&t_m(u+\tau)=(-1)^{mN}\,e^{-2i\pi\{mN(u+\frac{w}{n}+\frac{\tau}{2}-\frac{m-1}{2}w)-m\sum_{l=1}^N\theta_l\}}\,t_m(u),
\quad m=1,\cdots,n.\label{Quasi-transfer-4}
\eea

Let us evaluate the transfer matrix of the closed chain at some special points.
The initial condition of the $R$-matrix (\ref{Belavin-R}) implies that
\bea
t(\theta_j)=R_{j\,j-1}(\theta_j-\theta_{j-1})\cdots R_{j\,1}(\theta_j-\theta_{1})\,
            R_{j\,N}(\theta_j-\theta_{N})\cdots R_{j\,j+1}(\theta_j-\theta_{j+1}).\no
\eea
The unitarity relation (\ref{Unitarity}) allows us to derive the following identity:
\bea
\prod_{l=1}^N\,t(\theta_l)=\prod_{l=1}^N\,a(\theta_l)\,\times {\rm id},\quad a(u)=\prod_{l=1}^N\frac{\s(u-\theta_l+w)}{\s(w)},\quad d(u)=a(u-w).\label{product-transfer}
\eea

\subsection{Functional relations of eigenvalues}
The commutativity (\ref{commutive}) of the transfer matrices $\{t_m(u)|m=1,\cdots,n\}$  with different spectral parameters implies that they have common eigenstates.
Let $|\Psi\rangle$ be a common eigenstate of $\{t_m(u)\}$, which dose not depend upon $u$, with the eigenvalues $\Lambda_m(u)$ (we shall take the convention: $\L(u)=\L_1(u)$),
\bea
t_m(u)|\Psi\rangle=\Lambda_m(u)|\Psi\rangle,\qquad m=1,\cdots n.\no
\eea
The properties (\ref{quant-deter}), (\ref{Oper-1}) and (\ref{Zeros-Op}) of the transfer matrices $\{t_m(u)|m=1,\cdots,n\}$ imply that the corresponding eigenvalues
$\{\L_m(u)|m=1,\cdots,n\}$ satisfy the functional relations
\bea
&&\L(\theta_j)\,\L_m(\theta_j-w)=\L_{m+1}(\theta_j),\quad m=1,\cdots,n-1;\quad j=1,\cdots,N,\label{Eigenvalue-1}\\[4pt]
&&\L_m(\theta_j+kw)=0,\quad \quad j=1,\cdots,N;\,\, k=1,\cdots,m-1;\,\,m=2,\cdots,n-1,\label{Eigenvalue-2}\\[4pt]
&&\L_n(u)={\rm Det}_q\lt(T(u)\rt)=a(u)\,\prod_{k=1}^{n-1}\,d(u-kw),\label{Eigenvalue-3}
\eea
where the functions $a(u)$ and $d(u)$ are given by (\ref{product-transfer}). From the definitions (\ref{Belavin-R}), (\ref{transfer})
and (\ref{transfer-fused}) of the $R$-matrix $R(u)$ and the associated transfer matrices $\{t_m(u)|m=1,\cdots,n\}$, we have that
\bea
\hspace{-0.26truecm}\L_m\hspace{-0.08truecm}(u) \mbox{, as a function of $u$,
is an elliptical  polynomial of degree $mN$},\, m=1,\cdots,n\hspace{-0.12truecm}-\hspace{-0.12truecm}1.\label{Eigenvalue-4}
\eea
The periodicity (\ref{Quasi-transfer-3})-(\ref{Quasi-transfer-4}) of these transfer matrices imply that the eigenvalues $\{\L_m(u)\}$ are
some elliptic polynomials of the fixed degrees (\ref{Eigenvalue-4}) with the periodicity
\bea
\hspace{-1.2truecm}&&\L_m(u+1)=(-1)^{mN} \,\L_m(u),\quad m=1,\cdots,n-1,\label{Eigenvalue-5}\\[4pt]
\hspace{-1.2truecm}&&\L_m(u+\tau)=(-1)^{mN}\,e^{-2i\pi\{mN(u+\frac{w}{n}+\frac{\tau}{2}-\frac{m-1}{2}w)-m\sum_{l=1}^N\theta_l\}}\,\L_m(u),
\quad m=1,\cdots,n-1.\label{Eigenvalue-6}
\eea
Moreover the product identity (\ref{product-transfer}) of the transfer matrix $t(u)$ leads to the relation
\bea
 \prod_{l=1}^N\,\L(\theta_l)=\prod_{l=1}^N\,a(\theta_l),\label{Eigenvalue-7}
\eea  which serves as the selection rule \cite{Cao14} for the eigenvalues of the transfer matrix from the solutions of (\ref{Eigenvalue-1})-(\ref{Eigenvalue-6})

The relations (\ref{Eigenvalue-1})-(\ref{Eigenvalue-7}) allow us to determine the eigenvalues
$\{\L_m(u)\}$ of the transfer matrices $\{t_m(u)\}$ completely.


\section{ODBA solution of the $Z_3$ case}
\setcounter{equation}{0}
Similarly as that \cite{Cao14} for the eight-vertex model (the $Z_2$-case) ,  we  demonstrate
that (\ref{Eigenvalue-1})-(\ref{Eigenvalue-7}) enable us to express the eigenvalues
$\{\L_m(u)\}$ of the transfer matrices $\{t_m(u)\}$  simultaneously  in terms of some inhomogeneous $T-Q$ relations \cite{Wan15}.

For the $Z_3$-Belavin model, the corresponding (\ref{Eigenvalue-1})-(\ref{Eigenvalue-7}) read
\bea
\hspace{-1.2truecm}&&\Lambda(\theta_j)\Lambda_m(\theta_j-w)=\Lambda_{m+1}(\theta_j),~~~~~ m=1,2,\quad j=1,\cdots,N,\label{3-Eigenvalue-1}\\[4pt]
\hspace{-1.2truecm}&&\Lambda_3(u)=a(u)d(u-w)d(u-2w),\label{3-Eigenvalue-function-2}\\[4pt]
\hspace{-1.2truecm}&&\Lambda_2(\theta_j+w)=0,~~~~~~j=1,\cdots,N.\label{3-Eigenvalue-function-3} \\[4pt]
\hspace{-1.2truecm}&&\L_m(u+1)=(-1)^{mN} \,\L_m(u),\quad m=1,2,\label{3-Eigenvalue-4}\\[4pt]
\hspace{-1.2truecm}&&\L_m(u+\tau)=(-1)^{mN}\,e^{-2i\pi\{mN(u+\frac{w}{3}+\frac{\tau}{2}-\frac{m-1}{2}w)-m\sum_{l=1}^N\theta_l\}}\,\L_m(u),
\quad m=1,2.\label{3-Eigenvalue-5}\\[4pt]
\hspace{-1.2truecm}&&\prod_{l=1}^N\,\L(\theta_l)=\prod_{l=1}^N\,a(\theta_l).\label{3-Eigenvalue-7}
\eea
Keeping the fact that $\L_m(u)$ is an elliptical polynomial of degree $mN$ in mind, we can express
$\L(u)$ and $\L_2(u)$ in terms of the inhomogeneous $T-Q$ relation \cite{Cao14} as follows. Let us introduce some $Q$-functions
\bea
Q^{(i)}(u)=\prod_{j=1}^{N}\frac{\s(u-\l^{(i)}_j)}{\s(w)},\quad i=1,\cdots,4,
\eea parameterized by $4N$ parameters $\{\l^{(i)}_j|j=1,\cdots,N;i=1,\cdots,4\}$ (the so-called Bethe roots) determined later by the associated
BAEs (see below (\ref{3-BAE-1})-(\ref{3-BAE-9})). Associated with the
above $Q$-functions, we introduce 5 functions $\{Z_i(u)|i=1,2,3\}$ and $\{X_i(u)|i=1,2\}$ as
\bea
&&Z_1(u)=a(u)e^{2i\pi l_1 u+\phi_1}\frac{Q^{(1)}(u-w)}{Q^{(2)}(u)},\\[4pt]
&&Z_2(u)=d(u)e^{2i\pi l_2 u+\phi_2}\frac{Q^{(2)}(u+w)Q^{(3)}(u-w)}{Q^{(1)}(u)Q^{(4)}(u)},\\[4pt]
&&Z_3(u)=d(u)e^{-2i\pi\{(l_1+l_2)u+(2l_1+l_2)w\}-\phi_1-\phi_2}\frac{Q^{(4)}(u+w)}{Q^{(3)}(u)},\\[4pt]
&&X_1(u)=c_1a(u)d(u)e^{2i\pi l_3 u}\frac{Q^{(3)}(u-w)}{Q^{(1)}(u)Q^{(2)}(u)},\\[4pt]
&&X_2(u)=c_2a(u)d(u)e^{2i\pi l_4 u}\frac{Q^{(2)}(u+w)}{Q^{(3)}(u)Q^{(4)}(u)},
\eea
where $\{l_i|i=1,\cdots,4\}$ are 4 integers, $\{\phi_i,\,c_i|i=1,2\}$ are 4 complex numbers. Then we can introduce
the  inhomogeneous $T-Q$ relations,
\bea
&&\L(u)=Z_1(u)+Z_2(u)+Z_3(u)+X_1(u)+X_2(u), \label{3-T-Q-1}\\[4pt]
&&\L_2(u)=Z_1(u)Z_2(u-w)+Z_1(u)Z_3(u-w)+Z_2(u)Z_3(u-w)\no\\[4pt]
&&\qquad\qquad+X_1(u)Z_3(u-w)+Z_1(u)X_2(u-w).\label{3-T-Q-2}
\eea
In order that the above parameterizations of $\L(u)$ and $\L_2(u)$ become a solution to
(\ref{3-Eigenvalue-1})-(\ref{3-Eigenvalue-7}), the $4(N+1)$ parameters $\{\l^{(i)}_j|j=1,\cdots,N; i=1,\cdots,4\}$
and $\{\phi_i,\,c_i|i=1,2\}$ have to satisfy the associated BAEs
\bea
&&\hspace{-1.2truecm}e^{2i\pi l_2\l^{(1)}_j+\phi_2}Q^{(2)}(\l^{(1)}_j\hspace{-0.04truecm}+\hspace{-0.04truecm}w)Q^{(2)}(\l^{(1)}_j)
\hspace{-0.068truecm}+\hspace{-0.068truecm}c_1e^{2i\pi l_3\l^{(1)}_j}a(\l^{(1)}_j)Q^{(4)}(\l^{(1)}_j)
\hspace{-0.06truecm}=\hspace{-0.06truecm}0,\,j\hspace{-0.06truecm}=\hspace{-0.06truecm}1,\cdots,N,\label{3-BAE-1}\\[4pt]
&&\hspace{-1.2truecm}e^{2i\pi l_1\l^{(2)}_j+\phi_1}Q^{(1)}(\l^{(2)}_j\hspace{-0.04truecm}-\hspace{-0.04truecm}w)Q^{(1)}(\l^{(2)}_j)
\hspace{-0.068truecm}+\hspace{-0.068truecm}c_1e^{2i\pi l_3\l^{(2)}_j}d(\l^{(2)}_j)Q^{(3)}(\l^{(2)}_j\hspace{-0.04truecm}-\hspace{-0.04truecm}w)
\hspace{-0.06truecm}=\hspace{-0.06truecm}0,\,j\hspace{-0.06truecm}=\hspace{-0.06truecm}1,\cdots,N,\label{3-BAE-2}\\[4pt]
&&\hspace{-1.2truecm}e^{-2i\pi\{(l_1+l_2)\l^{(3)}_j+(2l_1+l_2)w\}-\phi_1-\phi_2}Q^{(4)}(\l^{(3)}_j\hspace{-0.04truecm}+\hspace{-0.04truecm}w)
Q^{(4)}(\l^{(3)}_j)\hspace{-0.068truecm}+\hspace{-0.068truecm}c_2e^{2i\pi l_4\l^{(3)}_j}a(\l^{(3)}_j)Q^{(2)}(\l^{(3)}_j\hspace{-0.04truecm}+\hspace{-0.04truecm}w)
\hspace{-0.06truecm}=\hspace{-0.06truecm}0,\no\\[4pt]
&&\qquad\qquad j=1,\cdots,N,\label{3-BAE-3}\\[4pt]
&&\hspace{-1.2truecm}e^{2i\pi l_2\l^{(4)}_j+\phi_2}Q^{(3)}(\l^{(4)}_j\hspace{-0.04truecm}-\hspace{-0.04truecm}w)Q^{(3)}(\l^{(4)}_j)
\hspace{-0.068truecm}+\hspace{-0.068truecm}c_2e^{2i\pi l_4\l^{(4)}_j}a(\l^{(4)}_j)Q^{(1)}(\l^{(4)}_j)
\hspace{-0.06truecm}=\hspace{-0.06truecm}0,\,j\hspace{-0.06truecm}=\hspace{-0.06truecm}1,\cdots,N,\label{3-BAE-4}
\eea
\bea
&&-\Theta^{(1)}+\Theta^{(2)}-\frac{N}{3}w=m_1+l_1\tau,\label{3-BAE-5}\\[4pt]
&&-\Theta^{(2)}-\Theta^{(3)}+\Theta^{(1)}+\Theta^{(4)}-\frac{N}{3}w=m_2+l_2\tau,\label{3-BAE-6}\\[4pt]
&&-\Theta-\Theta^{(3)}+\Theta^{(1)}+\Theta^{(2)}-\frac{N}{3}w=m_3+l_3\tau,\label{3-BAE-7}\\[4pt]
&&-\Theta-\Theta^{(2)}+\Theta^{(3)}+\Theta^{(4)}+\frac{5N}{3}w=m_4+l_4\tau,\label{3-BAE-8}\\[4pt]
&&\prod_{j=1}^N\frac{Q^{(1)}(\theta_j-w)}{Q^{(2)}(\theta_j)}=e^{-2i\pi l_1 \Theta-N\phi_1},\label{3-BAE-9}
\eea
where $\{m_i|i=1,\cdots,4\}$ are 4 integers and
\bea
\Theta=\sum_{l=1}^N\theta_l,\quad \Theta^{(i)}=\sum_{l=1}^N\l^{(i)}_l,\quad i=1,\cdots,4.
\eea

We have checked that the functions $\L(u)$ and $\L_2(u)$ given by the inhomogeneous $T-Q$ relations (\ref{3-T-Q-1})-
(\ref{3-T-Q-2}) are solutions to (\ref{3-Eigenvalue-1})-(\ref{3-Eigenvalue-7}) provided that the $4(N+1)$ parameters
$\{\l^{(i)}_j|j=1,\cdots,N;\,i=1,\cdots,4\}$ and $\phi_1$, $\phi_2$, $c_1$ and $c_2$ satisfy the associated BAEs (\ref{3-BAE-1})-
(\ref{3-BAE-9}) for arbitrary fixed integers $\{l_i,\,m_i|i=1,\cdots,4\}$. Therefore the corresponding $\L(u)$ becomes an eigenvalue of the
transfer matrix $t(u)$ given by (\ref{transfer}). In the homogeneous limit: $\{\theta_j\rightarrow 0\}$, the resulting $T-Q$ relation (\ref{3-T-Q-1}) and the associated
BAEs (\ref{3-BAE-1})-(\ref{3-BAE-9}) give rise to the eigenvalue and BAEs of the corresponding homogeneous spin chain
(i.e., the $Z_3$-Belavin model with periodic boundary condition described by the Hamiltonian (\ref{Ham}) for the case of $n=3$).

Some remarks are in order. The integers $\{l_i,\,m_i|i=1,\cdots,4\}$ appeared in the BAEs (\ref{3-BAE-5})-(\ref{3-BAE-8})
are due to the quasi-periodicity (\ref{Quasi-R-1})-(\ref{Quasi-R-2}) of the $R$-matrix in terms of the spectral parameter $u$.
Any choice of these integers may give rise to the complete set of eigenvalues $\L(u)$.  
Numerical solutions of the BAEs (\ref{3-BAE-1})-(\ref{3-BAE-9})  with random choice of $w$ and $\tau$ for some small size imply that the solution (\ref{3-T-Q-1}) indeed gives the complete solutions of the model. Here we present the numerical solutions of the BAEs for the $N=2$ case in Table \ref{num-N2};
The eigenvalue calculated from (\ref{3-T-Q-1}) is the same as that from the exact diagonalization of the Hamiltonian (\ref{Ham})  with periodic boundary condition  (\ref{Periodic-boundary}).
Moreover, for a generic $w$ and an arbitrary site number $N$, the eigenvalue
$\L(u)$ should be given by an inhomogeneous $T-Q$ relation such as (\ref{3-T-Q-1}) with non-vanishing terms related to $X_i(u)$. However,
when $N$ is some particular number (i.e., $N=3l$ for a positive integer $l$) or the crossing parameter $w$ takes some particular
values (i.e., see below (\ref{Discret-w-1})-(\ref{Discret-w-2})), the relation (\ref{3-T-Q-1}) is reduced to a homogeneous
one \cite{Bax82}, which corresponds to the $c_1=c_2=0$ solutions of (\ref{3-BAE-1})-(\ref{3-BAE-9}).


\begin{landscape}

\begin{table}
\caption{Solutions of BAEs (\ref{3-BAE-1})-(\ref{3-BAE-9}) for the $Z_3$ case, $N=2$, $\{\theta_j\}=0$, $w=-0.5$, $\tau=i$ and the parameters $m_1=1$, $m_2=m_3=m_4=l_1=l_2=l_3=l_4=0$. The symbol $m$ indicates the number of the eigenenergy $E$. }\label{num-N2}
\begin{tabular}{|c|c|c|c|c|c|c|}
\hline\hline
$\lambda^{(1)}_1$ & $\lambda^{(1)}_2$ & $\lambda^{(2)}_1$ & $\lambda^{(2)}_2$ & $\lambda^{(3)}_1$ \\ \hline
$-1.5000-0.2862i$  &  $1.5000+0.2862i$  &  $-0.1667+0.1501i$  &  $0.8333-0.1501i$  &  $-0.3053-1.0000i$ \\
$-1.5000+0.2862i$  &  $1.5000-0.2862i$  &  $-0.1667-0.1501i$  &  $0.8333+0.1501i$  &  $-0.3053+1.0000i$ \\
$0.2550+0.0000i$  &  $-0.2550-0.0000i$  &  $-0.1667+0.1501i$  &  $0.8333-0.1501i$  &  $0.7205-0.5000i$ \\
$-0.2795+0.5000i$  &  $0.2795-0.5000i$  &  $0.5633-0.5000i$  &  $0.1034+0.5000i$  &  $0.2550+1.0000i$ \\
$-0.2795-0.5000i$  &  $0.2795+0.5000i$  &  $0.5633+0.5000i$  &  $0.1034-0.5000i$  &  $0.2550-1.0000i$ \\
$0.7601-0.5000i$  &  $-0.7601+0.5000i$  &  $0.5633-0.5000i$  &  $0.1034+0.5000i$  &  $1.7517-0.0000i$ \\
$0.3053-0.0000i$  &  $-0.3053+0.0000i$  &  $1.5000-0.0000i$  &  $-0.8333+0.0000i$  &  $-0.5000-0.2862i$ \\
$0.3053-0.0000i$  &  $-0.3053+0.0000i$  &  $1.5000+0.0000i$  &  $-0.8333-0.0000i$  &  $-0.5000+0.2862i$ \\
$1.3053-1.0000i$  &  $-1.3053+1.0000i$  &  $0.1667+0.0000i$  &  $0.5000-0.0000i$  &  $0.5000-0.2862i$ \\
\hline\hline
$\lambda^{(3)}_2$ & $\lambda^{(4)}_1$ & $\lambda^{(4)}_2$ & $\phi_1$ & $\phi_2$ \\ \hline
$1.3053+1.0000i$  & $0.3333-0.0000i$ & $1.0000+0.0000i$  &  $-0.7466+6.2832i$  &  $-5.1156-5.5878i$ \\
$1.3053-1.0000i$  & $0.3333+0.0000i$ & $1.0000-0.0000i$  &  $-0.7466-6.2832i$  &  $-5.1156+5.5878i$ \\
$0.2795+0.5000i$  &  $1.3966-0.5000i$  &  $-0.0633+0.5000i$  &  $0.0054+3.1416i$  &  $0.0320-6.2239i$ \\
$0.7450-1.0000i$  & $0.6667+0.8499i$ & $0.6667-0.8499i$  &  $0.0374-0.3110i$  &  $-1.9185-5.9091i$ \\
$0.7450+1.0000i$  & $0.6667-0.8499i$ & $0.6667+0.8499i$  &  $0.0374+0.3110i$  &  $-1.9185+5.9091i$ \\
$-0.7517+0.0000i$  &  $-0.6667+0.0000i$  &  $2.0000-0.0000i$  &  $-0.0055-15.8970i$  &  $-0.0035+12.7554i$ \\
$1.5000+0.2862i$  & $1.6667+0.1501i$ & $-0.3333-0.1501i$  &  $0.4211-3.1416i$  &  $-1.1676+3.1416i$ \\
$1.5000-0.2862i$  & $1.6667-0.1501i$ & $-0.3333+0.1501i$  &  $0.4211+3.1416i$  &  $-1.1676-3.1416i$ \\
$0.5000+0.2862i$  &  $1.6667+0.1501i$  &  $-0.3333-0.1501i$  &  $-5.8621-0.6954i$  &  $5.1156+0.6954i$ \\
\hline\hline
$c_1$ & $c_2$ & $E$ & $m$ & \\ \hline
$-0.0069-0.0057i$  &  $-138.0911+115.2344i$ & $-5.619523$ & $1$ & \\
$-0.0069+0.0057i$  &  $-138.0911-115.2344i$ & $-5.619523$ & $1$ & \\
$0.0243-0.1299i$  &  $-2.7217+0.8747i$ & $-5.619523$ & $1$ & \\
$-0.0053-0.0007i$  &  $-346.9166+5524.2128i$ & $0.157726$ & $2$ & \\
$-0.0053+0.0007i$  &  $-346.9166-5524.2128i$ & $0.157726$ & $2$ & \\
$-2.8680+0.1801i$  &  $-0.0054+0.0427i$ & $0.157726$ & $2$ & \\
$-0.1592+0.0000i$  &  $3.1476+0.0000i$ & $5.461797$ & $3$ & \\
$-0.1592+0.0000i$  &  $3.1476+0.0000i$ & $5.461797$ & $3$ & \\
$65.4540+54.6201i$  &  $3.1476-0.0000i$ & $5.461797$ & $3$ & \\
\hline
\end{tabular}
\end{table}
\end{landscape}

\subsection{Generic $w$ and $\tau$ case}

It follows from (\ref{3-BAE-1})-(\ref{3-BAE-4}) that for the solution with $c_1=c_2=0$, the parameters
$\{\l^{(i)}_j\}$ have to form the pairs:
\bea
\lt\{\begin{array}{ll}\l^{(1)}_j=\l_k^{(2)},&{\rm or}\,\,\l^{(1)}_j=\l_k^{(2)}-w.\\[6pt]
\l^{(3)}_j=\l_k^{(4)},&{\rm or}\,\,\l^{(3)}_j=\l_k^{(4)}-w \end{array}\rt..
\eea
Without losing generality, let us suppose that
\bea
&&\l^{(1)}_j= \l^{(2)}_j\stackrel{{\rm Redef}}{=}\bar{\l}^{(1)}_j,\quad j=1,\cdots,\bar{M}_1,\no\\[6pt]
&&\l^{(1)}_{\bar{M}_1+j}= \l^{(2)}_{\bar{M}_1+j}-w\stackrel{{\rm Redef}}{=}\bar{\l}^{(1)}_{\bar{M}_1+j},\quad j=1,\cdots,N-\bar{M}_1,\no\\[6pt]
&&\l^{(3)}_j= \l^{(4)}_j\stackrel{{\rm Redef}}{=}\bar{\l}^{(2)}_j,\quad j=1,\cdots,\bar{M}_2,\no\\[6pt]
&&\l^{(3)}_{\bar{M}_2+j}= \l^{(4)}_{\bar{M}_2+j}-w\stackrel{{\rm Redef}}{=}\bar{\l}^{(2)}_{\bar{M}_2+j},\quad j=1,\cdots,N-\bar{M}_2,\no
\eea where $\bar{M}_1$ and $\bar{M}_2$ are two non-negative integers.
The corresponding $T-Q$ relation (\ref{3-T-Q-1}) is reduced to
\bea
\hspace{-1.2truecm}\L(u)&=&a(u)e^{2i\pi l_1 u+\phi_1}\frac{\bar{Q}^{(1)}(u-w)}{\bar{Q}^{(1)}(u)}+
d(u)e^{2i\pi l_2 u+\phi_2}\frac{\bar{Q}^{(1)}(u+w)\bar{Q}^{(2)}(u-w)}{\bar{Q}^{(1)}(u)\bar{Q}^{(2)}(u)}\no\\[6pt]
\hspace{-1.2truecm}&&\quad +d(u)e^{-2i\pi\{(l_1+l_2)u+(2l_1+l_2)w\}-\phi_1-\phi_2}\frac{\bar{Q}^{(2)}(u+w)}{\bar{Q}^{(2)}(u)},\label{3-T-Q-3}
\eea
where the reduced $Q$-functions are given by
\bea
\bar{Q}^{(i)}(u)=\prod_{j=1}^{\bar{M}_i}\frac{\s(u-\bar{\l}^{(i)}_j)}{\s(w)},\quad i=1,2,
\eea provided that the two non-negative integers $\bar{M}_1$ and $\bar{M}_2$ satisfy the relations
\bea
\lt\{\begin{array}{l}(\frac{2}{3}N-\bar{M}_1)w=m_1+l_1\tau\\[6pt]
(\bar{M}_1-\bar{M}_2-\frac{N}{3})w=m_2+l_2\tau
\end{array}\rt.,\label{Reduced-Condition}
\eea
where $m_1$, $m_2$, $l_1$ and $l_2$ are some integers.

\begin{itemize}
\item For the case of $N=3l$ with a positive integer $l$. The only solution to (\ref{Reduced-Condition}) is
\bea
m_1=m_2=l_1=l_2=0,\quad {\rm and}\,\, \bar{M}_1=2l,\,\bar{M}_2=l. \no
\eea
The resulting $T-Q$ relation becomes
\bea
\L(u)&=&a(u)e^{\phi_1}\frac{\bar{Q}^{(1)}(u-w)}{\bar{Q}^{(1)}(u)}+
d(u)e^{\phi_2}\frac{\bar{Q}^{(1)}(u+w)\bar{Q}^{(2)}(u-w)}{\bar{Q}^{(1)}(u)\bar{Q}^{(2)}(u)}\no\\[6pt]
&&\quad +d(u)e^{-\phi_1-\phi_2}\frac{\bar{Q}^{(2)}(u+w)}{\bar{Q}^{(2)}(u)},\label{3-T-Q-3-R}
\eea
the $3l$ parameters $\{\bar{\l}^{(1)}_j|j=1,\cdots,2l\}$ and $\{\bar{\l}^{(2)}_j|j=1,\cdots,l\}$ satisfy the associated
BAEs and the selection rule
\bea
&&\frac{a(\bar{\l}^{(1)}_j)\,\bar{Q}^{(2)}(\bar{\l}^{(1)}_j)\,e^{\phi_1-\phi_2}}{d(\bar{\l}^{(1)}_j)\,\bar{Q}^{(2)}(\bar{\l}^{(1)}_j-w)}
= -\frac{\bar{Q}^{(1)}(\bar{\l}^{(1)}_j+w)}{\bar{Q}^{(1)}(\bar{\l}^{(1)}_j-w)},\quad j=1,\cdots,2l,\\[6pt]
&&e^{\phi_1+2\phi_2}\frac{\bar{Q}^{(1)}(\bar{\l}^{(2)}_j+w)}{\bar{Q}^{(1)}(\bar{\l}^{(2)}_j)}
= -\frac{\bar{Q}^{(2)}(\bar{\l}^{(2)}_j+w)}{\bar{Q}^{(2)}(\bar{\l}^{(2)}_j-w)},\quad j=1,\cdots,l,\\[6pt]
&&\prod_{j=1}^N\frac{Q^{(1)}(\theta_j-w)}{Q^{(1)}(\theta_j)}=e^{-N\phi_1}.
\eea
For this case (i.e., $N=3l$), the algebraic Bethe Ansatz can also be applied to and our results recover
those obtained in \cite{Hou89,Hou03}.

\item $N\neq 3l$ case. Since $\tau$ and $w$ are generic complex numbers, generally (\ref{Reduced-Condition}) can not be satisfied in this case
 and the eigenvalue
$\L(u)$ should be given by an inhomogeneous $T-Q$ relation.

\end{itemize}


\subsection{Degenerate $w$ case}

For some degenerate values of $w$, $c_1=c_2=0$ solutions indeed exist for an arbitrary site number $N$. In this case, the parameters $w$ and $\tau$
are no longer independent but related with the  constraint condition:
\bea
w=\frac{3m_1+3l_1\tau}{2N-3\bar{M}_1},\quad {\rm for}\quad m_1,l_1\in \Zb;\quad\bar{M}_1\in\Zb^{+}, \label{Discret-w-1}
\eea
and there exists an integer $n_1$ such that
\bea
\bar{M}_2=(n_1+1)\bar{M}_1-\frac{2n_1+1}{3}N \in \Zb^{+}. \label{Discret-w-2}
\eea
In this case the relation (\ref{Reduced-Condition}) is fulfilled by
\bea
\lt\{\begin{array}{l}(\frac{2}{3}N-\bar{M}_1)w=m_1+l_1\tau\\[6pt]
(\bar{M}_1-\bar{M}_2-\frac{N}{3})w=n_1(m_1+l_1\tau)
\end{array}\rt..\no
\eea
The resulting $T-Q$ relation becomes
\bea
\L(u)&=&a(u)e^{2i\pi l_1 u+\phi_1}\frac{\bar{Q}^{(1)}(u-w)}{\bar{Q}^{(1)}(u)}+
d(u)e^{2i\pi n_1l_1 u+\phi_2}\frac{\bar{Q}^{(1)}(u+w)\bar{Q}^{(2)}(u-w)}{\bar{Q}^{(1)}(u)\bar{Q}^{(2)}(u)}\no\\[6pt]
&&\quad +d(u)e^{-2i\pi\{(n_1+1)l_1 u+(n_1+2)l_1 w\}-\phi_1-\phi_2}\frac{\bar{Q}^{(2)}(u+w)}{\bar{Q}^{(2)}(u)}.\label{3-T-Q-R-2}
\eea
The resulting BAEs and selection rule read
\bea
&&e^{2i\pi(1-n_1)l_1\l^{(1)}_j}\frac{a(\bar{\l}^{(1)}_j)\,\bar{Q}^{(2)}(\bar{\l}^{(1)}_j)\,e^{\phi_1-\phi_2}}{d(\bar{\l}^{(1)}_j)\,\bar{Q}^{(2)}(\bar{\l}^{(1)}_j-w)}
= -\frac{\bar{Q}^{(1)}(\bar{\l}^{(1)}_j+w)}{\bar{Q}^{(1)}(\bar{\l}^{(1)}_j-w)},\quad j=1,\cdots,\bar{M}_1,\\[6pt]
&&e^{2i\pi\{(2n_1+1)l_1 \l^{(2)}_j+(n_1+2)l_1 w\}+\phi_1+2\phi_2}\frac{\bar{Q}^{(1)}(\bar{\l}^{(2)}_j+w)}{\bar{Q}^{(1)}(\bar{\l}^{(2)}_j)}
= -\frac{\bar{Q}^{(2)}(\bar{\l}^{(2)}_j+w)}{\bar{Q}^{(2)}(\bar{\l}^{(2)}_j-w)},\no\\[4pt]
&&\quad\quad\quad\quad j=1,\cdots,\bar{M}_2,\\[6pt]
&&\prod_{j=1}^N\frac{Q^{(1)}(\theta_j-w)}{Q^{(1)}(\theta_j)}=e^{-N\phi_1-2i\pi l_1 \sum_{j=1}^N\theta_j}.
\eea


\section{Results for the $Z_n$ case}
\setcounter{equation}{0}

In Sections 3, we have obtained the very operator product identities (\ref{Oper-1})-(\ref{product-transfer}) for  the fused transfer matrices
$\{t_j(u)|j=1,\cdots, n\}$. These identities lead to that the corresponding eigenvalues $\{\L_j(u)|j=1,\cdots, n\}$ of the transfer matrices satisfy
the associated relations (\ref{Eigenvalue-1})-(\ref{Eigenvalue-7}). Similarly as those for the $Z_3$ case, the relations allow us to determine the eigenvalues of the transfer matrix of the $Z_n$-Belavin model  completely.

Let us
introduce some functions $\{Q^{(i)}|i=1,\cdots,2n-2\}$,
$\{Z_i|i=1,\cdots,n\}$ and $\{X_i|i=1,\cdots,n-1\}$ as follows:
\bea
&&Q^{(i)}(u)=\prod_{j=1}^{N_i}\frac{\s(u-\l^{(i)}_j)}{\s(w)},\quad i=1,\ldots,2n-2,\label{n-Q-function}\\[6pt]
&&Z_1(u)=e^{2i\pi{l_1}u+\phi_1}a(u)\frac{Q^{(1)}(u-w)}{Q^{(2)}(u)},\no\\[6pt]
&&Z_2(u)=e^{2i\pi{l_2}u+\phi_2}d(u)\frac{Q^{(2)}(u+w)Q^{(3)}(u-w)}{Q^{(1)}(u)Q^{(4)}(u)},\no\\[6pt]
&&\qquad\qquad\vdots\no\\[4pt]
&&Z_i(u)=e^{2i\pi{l_i}u+\phi_i} d(u)\frac{Q^{(2i-2)}(u+w)Q^{(2i-1)}(u-w)}{Q^{(2i-3)}(u)Q^{(2i)}(u)},\no\\[6pt]
&&\qquad\qquad\vdots\no\\[4pt]
&&Z_{n-1}(u)=e^{2i\pi{l_{n-1}}u+\phi_{n-1}} d(u)\frac{Q^{(2n-4)}(u+w)Q^{(2n-3)}(u-w)}{Q^{(2n-5)}(u)Q^{(2n-2)}(u)},\no\\[6pt]
&&Z_n(u)=e^{-2i\pi{\sum^{n-1}_{k=1}l_k(u+(n-k)w)-\sum_{j=1}^{n-1}\phi_j}}d(u)\frac{Q^{(2n-2)}(u+w)}{Q^{(2n-3)}(u)},
\label{n-homo}\eea
and
\bea
&&X_1(u)=c_1e^{2i\pi{l_n}u}a(u)d(u)\frac{Q^{(3)}(u-w)}{Q^{(1)}(u)Q^{(2)}(u)},\no\\[6pt]
&&X_2(u)=c_2e^{2i\pi{l_{n+1}}u} a(u)d(u)\frac{Q^{(2)}(u+w)Q^{(5)}(u-w)}{Q^{(3)}(u)Q^{(4)}(u)},\no\\[6pt]
&&\qquad\qquad\vdots\no\\[4pt]
&&X_j(u)=c_j e^{2i\pi{l_{n+j-1}}u} a(u)d(u)\frac{Q^{(2j-2)}(u+w)Q^{(2j+1)}(u-w)}{Q^{(2j-1)}(u)Q^{(2j)}(u)},\no\\[6pt]
&&\qquad\qquad\vdots\no\\[4pt]
&&X_{n-1}(u)=c_{n-1}e^{2i\pi{l_{2n-2}}u}a(u)d(u)\frac{Q^{(2n-4)}(u+w)}{Q^{(2n-3)}(u)Q^{(2n-2)}(u)},
\label{n-inhomo}\eea
where the $2n-2$ positive integers $\{N_i|i=1,\cdots,2n-2\}$ are given by (\ref{N-1})-(\ref{N-3}), $\{l_i|i=1,\cdots,2n-2\}$ are arbitrary integers,
$\{\phi_i,c_i|i=1,\cdots,n-1\}$ are $2n-2$ complex numbers.
It is remarked that for an even $n$, an extra factor function $f_{n\over2}(u)$ should be added to the function $X_{n\over2}(u)$, namely,
\bea
&&X_{n\over2}(u)=c_{n\over2} e^{2i\pi{l_{{3n\over2}-1}}u} a(u)d(u)\frac{Q^{(n-2)}(u+w)Q^{(n+1)}(u-w)}{Q^{(n-1)}(u)Q^{(n)}(u)}\times f_{n\over2}(u),\label{middleX}
\eea
which ensures that all the numbers $\{N_i|i=1,\cdots,2n-2\}$ are positive integers. The explicit expression of the function $f_{n\over2}(u)$ is given by (\ref{f-funtion-1-2}) (or (\ref{f-function-3})) below.

We are now in position to construct the associated inhomogeneous $T-Q$ relations similar to those given by (\ref{3-T-Q-1})-(\ref{3-T-Q-2}). Let us introduce  the  functions $\{Y_l(u)|l=1,\cdots,2n-1\}$,
\bea
\lt\{\begin{array}{ll}Y_{2j-1}(u)=Z_j(u),& j=1,\cdots,n,\\[6pt]
Y_{2j}(u)=X_j(u),& j=1,\cdots,n-1.
\end{array}\rt.\label{Def-Y}
\eea
We further take the notation:
\bea
Y_j^{(l)}(u)=Y_j(u-lw),\quad l=1,\cdots,n,\quad j=1,\cdots,2n-1.\label{Notation-1}
\eea
Then the eigenvalue  $\{\Lambda_m(u)|m=1,\cdots n-1\}$ which satisfy the relations (\ref{Eigenvalue-1})-(\ref{Eigenvalue-7}) can
be given in terms of the inhomogeneous $T-Q$ relations
\bea
\Lambda_m(u)={\sum}'_{1\leq i_1<i_2<\cdots<i_m\leq 2n-1}Y_{i_1}(u)Y^{(1)}_{i_2}(u)\cdots Y^{(m-1)}_{i_m}(u),\quad
m=1,\cdots,n-1.\label{T-Q-n-1}
\eea
The sum $\sum'$ in the above expression is over the constrained increasing sequences $1\leq i_1<i_2<\cdots<i_m\leq 2n-1$
such that when any $i_k=2j$ (i.e., $Y_{i_k}^{(k-1)}(u)=Y_{2j}^{(k-1)}(u)=X^{(k-1)}_j(u)$), $i_{k-1}\leq 2j-3$  and $i_{k+1}\geq 2j+3$.
Namely, when $Y_{i_k}(u)=Y_{2j}=X_j(u)$,  the previous element $Y_{i_{k-1}}(u)$ can not be chosen as $Y_{2j-1}=Z_j(u)$ or $Y_{2j-2}=X_{j-1}(u)$,
while the next element $Y_{i_{k+1}}(u)$ can not be chosen as $Y_{2j+1}=Z_{j+1}(u)$ or $Y_{2j+2}=X_{j+1}(u)$ (e.g.,
once a $X_j$ element was chosen, its nearest neighbors (namely, $X_{j-1}(u)$, $Z_j(u)$, $Z_{j+1}(u)$ and $X_{j+1}(u)$) in the diagram (\ref{struc-inhomo})
can not be chosen any more.).
\bea
\xymatrix{
Z_1\ar@{-}[rr]\ar@{-}[dr]&   &Z_2\ar@{-}[dr]\ar@{-}[rr]&   &Z_3\ar@{-}[dl]\ar@{-}[rr]&   &Z_4\ar@{-}[dl]\ar@{.}[rr]& &Z_{n}\ar@{.}[dl]\\
   &X_1\ar@{-}[rr] \ar@{-}[ur]&   &X_2\ar@{-}[rr]&   &X_3\ar@{-}[ul]\ar@{.}[rr]&   &X_{n-1}}
\label{struc-inhomo}\eea

The $2n-2$ positive integers $\{N_i|i=1,\cdots,2n-2\}$ and the function $f_{n\over2}(u)$ in (\ref{middleX}) are given as follows:
\begin{itemize}
\item For the case of odd $n$, we have
\bea
N_{2i-1}=N_{2i}=N_{2(n-i)-1}=N_{2(n-i)}=\frac{i(n-i)}{2}N,\quad
i=1,\cdots,\frac{n-1}{2};\label{N-1}
\eea
and there is no function $X_{n\over2}(u)$;
\item For the case of even $n$ and even $N$, we have
\bea
N_{2i-1}=N_{2i}=N_{2(n-i)-1}=N_{2(n-i)}=\frac{i(n-i)}{2}N,\quad
i=1,\cdots,\frac{n}{2};\label{N-2}
\eea
and
\bea
f_{n\over2}(u)=1; \label{f-funtion-1-2}
\eea
\item For the case of even $n$ and odd $N$, we have
\bea
N_{2i-1}=N_{2i}=N_{2(n-i)-1}=N_{2(n-i)}=\frac{i(n-i)}{2}N+\frac{i}{2},\quad
i=1,\cdots,\frac{n}{2};\label{N-3}
\eea
the function $f_{\frac{n}{2}}(u)$ is given by
\bea
f_{n\over2}(u)=\sigma(u).\label{f-function-3}
\eea

\end{itemize}
Moreover, the vanishing condition of the residues of $\L_m(u)$ at the points $\l^{(i)}_j$ gives rise to the BAEs:
\bea
\hspace{-0.8truecm}&&e^{2i\pi{l_2}\lambda^{(1)}_j+\phi_2}\frac{Q^{(2)}(\lambda^{(1)}_j+w)}{Q^{(4)}(\lambda^{(1)}_j)}
+c_1e^{2i\pi{l_n}\lambda^{(1)}_j}\frac{a(\lambda^{(1)}_j)}{Q^{(2)}(\lambda^{(1)}_j)}=0,\qquad j=1,\cdots,N_1,\label{BAE-1}\\[6pt]
\hspace{-0.8truecm}&&e^{2i\pi{l_1}\lambda^{(2)}_j+\phi_1}Q^{(1)}(\lambda^{(2)}_j-w)+
c_1e^{2i\pi{l_n}\lambda^{(2)}_j}d(\lambda^{(2)}_j)\frac{Q^{(3)}(\lambda^{(2)}_j-w)}{Q^{(1)}(\lambda^{(2)}_j)}=0,\quad j=1,\cdots,N_2,\\[6pt]
\hspace{-0.8truecm}&&\qquad\qquad\vdots\no\\[4pt]
\hspace{-0.8truecm}&&e^{2i\pi{l_{i+1}}\lambda^{(2i-1)}_j+\phi_{i+1}}\frac{Q^{(2i)}(\lambda^{(2i-1)}_j+w)}{Q^{(2i+2)}(\lambda^{(2i-1)}_j)}
\hspace{-1mm}+\hspace{-1mm}c_ie^{2i\pi{l_{n+i-1}}\lambda^{(2i-1)}_j}
\hspace{-1mm}a(\lambda^{(2i-1)}_j)\frac{Q^{(2i-2)}(\lambda^{(2i-1)}_j+w)}{Q^{(2i)}(\lambda^{(2i-1)}_j)}=0,\no\\[6pt]
\hspace{-0.8truecm}&&\qquad \qquad i=2,\cdots,n-2;j=1,\cdots,N_{2i-1},\\[6pt]
\hspace{-0.8truecm}&&e^{2i\pi{l_{i}}\lambda^{(2i)}_j+\phi_{i}}\frac{Q^{(2i-1)}(\lambda^{(2i)}_j-w)}{Q^{(2i-3)}(\lambda^{(2i)}_j)}
+c_ie^{2i\pi{l_{n+i-1}}\lambda^{(2i)}_j}a(\lambda^{(2i)}_j)\frac{Q^{(2i+1)}(\lambda^{(2i)}_j-w)}{Q^{(2i-1)}(\lambda^{(2i)}_j)}=0,\no\\[6pt]
\hspace{-0.8truecm}&&\qquad\qquad i=2,\cdots,n-2;j=1,\cdots,N_{2i},\\[6pt]
\hspace{-0.8truecm}&&\qquad\qquad\vdots\no\\[4pt]
\hspace{-0.8truecm}&&e^{-2i\pi{\sum^{n-1}_{k=1}l_k(\lambda^{(2n-3)}_j+(n-k)w)-\sum_{l=1}^{n-1}\phi_l}}Q^{(2n-2)}(\lambda^{(2n-3)}_j+w)\no\\[6pt]
\hspace{-0.8truecm}&&~~~~+c_{n-1}e^{2i\pi{l_{2n-2}}\lambda^{(2n-3)}_j}a(\lambda^{(2n-3)}_j)\frac{Q^{(2n-4)}(\lambda^{(2n-3)}_j+w)}{Q^{(2n-2)}(\lambda^{(2n-3)}_j)}=0,
\quad j=1,\cdots,N_{2n-3},\\[6pt]
\hspace{-0.8truecm}&&e^{2i\pi{l_{n-1}}\lambda^{(2n-2)}_j+\phi_{n-1}}\frac{Q^{(2n-3)}(\lambda^{(2n-2)}_j-w)}{Q^{(2n-5)}(\lambda^{(2n-2)}_j)}
+c_{n-1}e^{2i\pi{l_{2n-2}}\lambda^{(2n-2)}_j}\frac{a(\lambda^{(2n-2)}_j)}{Q^{(2n-3)}(\lambda^{(2n-2)}_j)}=0,\no\\[6pt]
\hspace{-0.8truecm}&&~~~~~~~~~~~j=1,\cdots,N_{2n-2}.
\eea
Further, the  periodicities (\ref{Eigenvalue-6}) of the eigenvalues  as well as the selection rule (\ref{Eigenvalue-7}) give rise to the associated BAEs:
\bea
\hspace{-1.2truecm}&&\Theta^{(2)}-\Theta^{(1)}=(N_1-N+\frac{N}{n})w+m_1+l_1\tau,\\[6pt]
\hspace{-1.2truecm}&&\Theta^{(1)}-\Theta^{(2)}-\Theta^{(3)}+\Theta^{(4)}=(N_3-N_2+\frac{N}{n})w+m_2+l_2\tau,\\[6pt]
\hspace{-1.2truecm}&&\qquad\qquad\vdots\no\\[4pt]
\hspace{-1.2truecm}&&-\Theta^{(2i-2)}-\Theta^{(2i-1)}+
\Theta^{(2i-3)}+
\Theta^{(2i)}=(N_{2i-1}-N_{2i-2}+\frac{N}{n})w+m_i+l_i\tau,\\[6pt]
\hspace{-1.2truecm}&&\qquad\qquad\vdots\no\\[4pt]
\hspace{-1.2truecm}&&-\Theta^{(2n-4)}\hspace{-1mm}-\Theta^{(2n-3)}\hspace{-1mm}+\Theta^{(2n-5)}\hspace{-1mm}+
\Theta^{(2n-2)}\hspace{-1mm}=(N_{2n-3}\hspace{-1mm}-N_{2n-4}\hspace{-1mm}+\hspace{-1mm}
\frac{N}{n})w+m_{n-1}\hspace{-1mm}+\hspace{-1mm}l_{n-1}\tau,\\[6pt]
\hspace{-1.2truecm}&&-\Theta+\Theta^{(1)}+\Theta^{(2)}-\Theta^{(3)}+\frac{(n-1)Nw}{n}-N_3w=l_n\tau+m_n,\\[6pt]
\hspace{-1.2truecm}&&\qquad\qquad\vdots\no\\[4pt]
\hspace{-1.2truecm}&&-\Theta-\Theta^{(2j-2)}+\Theta^{(2j-1)}+\Theta^{(2j)}-\Theta^{(2j+1)}+(N_{2j-2}-N_{2j+1}+\frac{(n-1)N}{n})w\no\\[6pt]
\hspace{-1.2truecm}&&~~~~~~~~~~~~~~~~~~~~~~~~~~~~~~~~~~~~~~~~~~~~~~~~~~=l_{n+j-1}\tau+m_{n+j-1},\\[6pt]
\hspace{-1.2truecm}&&\qquad\qquad\vdots\no\\[4pt]
\hspace{-1.2truecm}&&-\Theta-\Theta^{(2n-4)}+\Theta^{(2n-3)}+\Theta^{(2n-2)}+(N_{2n-4}+\frac{(n-1)N}{n})w=l_{2n-2}\tau+m_{2n-2},\\[6pt]
\hspace{-1.2truecm}&&\prod^{N}_{j=1}\frac{Q^{(1)}(\theta_j-w)}{Q^{(2)}(\theta_j)}=e^{-2i\pi{l_1}\Theta-N\phi_1},\label{n-BAE-prod}
\eea
where $\{m_i|i=1,\cdots,2n-2\}$ are arbitrary integers  and
\bea
\Theta=\sum_{l=1}^N\theta_l,\quad \Theta^{(i)}=\sum_{l=1}^{N_i}\l^{(i)}_l,\quad i=1,\cdots,2n-2.
\eea

We have checked that for a generic $w$ and $\tau$ but the number of sites $N=nl$ with $l$ being a positive integer, the inhomogeneous
$T-Q$ relations (\ref{T-Q-n-1}) can be reduced to homogeneous ones which were previously obtained by the algebraic Bethe ansatz \cite{Hou89,Hou03}.
Moreover, it is also found that when the crossing parameter $w$ takes some discrete values (like (\ref{Discret-w-1}) for the $n=3$ case) the resulting
$T-Q$ relations can also  become the homogeneous ones.


\section{Conclusions}
\setcounter{equation}{0}

The periodic $Z_n$-Belavin model with an arbitrary site number $N$ and generic coupling constants $w$ and $\tau$
described by the Hamiltonian (\ref{Ham}) and (\ref{Periodic-boundary}) is studied via the off-diagonal Bethe
Ansatz  method. The eigenvalues $\{\L_i(u)|i=1,\cdots,n-1\}$ of the corresponding transfer matrix and fused ones $\{t_i(u)|i=1,\cdots,n-1\}$ given by
(\ref{transfer-fused})  are derived in terms of the inhomogeneous $T-Q$ relations (\ref{T-Q-n-1}). In the special
case of $N=nl$ with a positive integer $l$, the resulting $T-Q$ relation is reduced to a
homogeneous one (such as (\ref{3-T-Q-3-R})), which recovers the result  obtained by the algebraic Bethe Ansatz method \cite{Hou89}.
On the other hand, if the crossing parameter $w$ take some special values (such as (\ref{Discret-w-1}) for the $n=3$ case),
the resulting $T-Q$ relation also becomes a homogeneous one (such as (\ref{3-T-Q-R-2}) for the $n=3$ case).

We remark that the $Z_n$-symmetry (\ref{Zn-symmetry}) of the $R$-matrix $R(u)$ ensures that the $Z_n$-Belavin model with the twisted boundary condition given by
\bea
H_{N,N+1}=G_1\,H_{N,1}\,G_1^{-1},\quad G=I_{{\bf \alpha}}=I_{(\a_1,\a_2)},\quad \a_i\in\Zb_n,\label{Twisted-condition}
\eea
is also integrable. The corresponding transfer matrix $t^{({\bf \a})}(u)$ can be constructed by \cite{deV84,Bat95}
\bea
t^{({\bf \a})}(u)= tr_0\lt(G_0\,T_0(u)\rt),\quad G=I_{{\bf \alpha}}=I_{(\a_1,\a_2)},\quad \a_i\in\Zb_n.\label{Transfer-twisted}
\eea
The Hamiltonian can be derived the same way as the periodic one (c.f., (\ref{Ham})). Using the similar method developed in previous sections,
we can construct the corresponding ODBA solution, which  is given in Appendix B.

The eigenvalues of the transfer matrix for the $Z_n$-Belavin model with periodic (or twisted) boundary condition obtained in this paper might help
one to construct the corresponding eigenstates, thus further giving rise to studying correlation functions \cite{Kor93} of the model. For this purpose,
some particular basis such as the separation of variable (SoV) \cite{Skl85} basis \cite{Nic13, Nic15} or its higher-rank generalization \cite{Hao16} will
play an important role.

\section*{Acknowledgments}

The financial supports from the National Natural Science Foundation
of China (Grant Nos. 11375141, 11374334, 11434013, 11425522 and
11547045), BCMIIS,  the National Program for Basic Research of
MOST (Grant No. 2016YFA0300603) and the Strategic Priority
Research Program of the Chinese Academy of Sciences are gratefully
acknowledged. Two of the authors (K. Hao and F. Wen) would like to thank IoP/CAS for the hospitality
during their visit there. They also would like to acknowledge S. Cui for his numerical helps.

\section*{Appendix A: $T-Q$ relations for the $Z_4$ case}
\setcounter{equation}{0}
\renewcommand{\theequation}{A.\arabic{equation}}

In this Appendix, we take the $Z_4$ case as an example to show the procedure for constructing the inhomogeneous $T-Q$ relations (\ref{T-Q-n-1}). The functions (\ref{n-Q-function})-(\ref{n-inhomo}) now read
\bea
&&Q^{(i)}(u)=\prod_{j=1}^{N_i}\frac{\s(u-\l^{(i)}_j)}{\s(w)},\quad i=1,\cdots,6,\\[6pt]
&&Z_1(u)=e^{2i\pi{l_1}u+\phi_1}a(u)\frac{Q^{(1)}(u-w)}{Q^{(2)}(u)},\no\\[6pt]
&&Z_2(u)=e^{2i\pi{l_2}u+\phi_2}d(u)\frac{Q^{(2)}(u+w)Q^{(3)}(u-w)}{Q^{(1)}(u)Q^{(4)}(u)},\no\\[6pt]
&&Z_3(u)=e^{2i\pi{l_3}u+\phi_3} d(u)\frac{Q^{(4)}(u+w)Q^{(5)}(u-w)}{Q^{(3)}(u)Q^{(6)}(u)},\no\\[6pt]
&&Z_4(u)=e^{-2i\pi{\sum^{3}_{k=1}l_k(u+(4-k)w)-\sum_{j=1}^{3}\phi_j}}d(u)\frac{Q^{(6)}(u+w)}{Q^{(5)}(u)},
\eea

and
\bea
&&X_1(u)=c_1e^{2i\pi{l_4}u}a(u)d(u)\frac{Q^{(3)}(u-w)}{Q^{(1)}(u)Q^{(2)}(u)},\no\\[6pt]
&&X_2(u)=c_2e^{2i\pi{l_5}u} a(u)d(u)\frac{Q^{(2)}(u+w)Q^{(5)}(u-w)f_2(u)}{Q^{(3)}(u)Q^{(4)}(u)},\no\\[6pt]
&&X_3(u)=c_3e^{2i\pi{l_6}u}a(u)d(u)\frac{Q^{(4)}(u+w)}{Q^{(5)}(u)Q^{(6)}(u)}.
\eea
The inhomogeneous $T-Q$ relations (\ref{T-Q-n-1}) become
\bea
&&\hspace{-1.6truecm}\L(u)=Z_1(u)+Z_2(u)+Z_3(u)+Z_4(u)+X_1(u)+X_2(u)+X_3(u),\\[6pt]
&&\hspace{-1.6truecm}\L_2(u)=Z_1(u)Z_2(u-w)+Z_1(u)Z_3(u-w)+Z_1(u)Z_4(u-w)+Z_2(u)Z_3(u-w)\no\\[6pt]
&&+Z_2(u)Z_4(u-w)+Z_3(u)Z_4(u-w)
+X_1(u)(Z_3(u-w)+Z_4(u-w))\no\\[6pt]
&&+(Z_1(u)+Z_2(u))X_3(u-w)+X_1(u)X_3(u-w)+Z_1(u)X_2(u-w)\no\\[6pt]
&&+X_2(u)Z_4(u-w),\\[6pt]
&&\hspace{-1.6truecm}\L_3(u)=Z_1(u)Z_2(u-w)Z_3(u-2w)+Z_1(u)Z_2(u-w)Z_4(u-2w)\no\\[6pt]
&&+Z_1(u)Z_3(u-w)Z_4(u-2w)+Z_2(u)Z_3(u-w)Z_4(u-2w)\no\\[6pt]
&&+Z_1(u)Z_2(u-w)X_3(u-2w)+Z_1(u)X_2(u-w)Z_4(u-2w)\no\\[6pt]
&&+X_1(u)Z_3(u-w)Z_4(u-2w),\\[6pt]
&&\hspace{-1.6truecm}\Lambda_4(u)=Z_1(u)Z_2(u-w)Z_3(u-2w)Z_4(u-3w).
\eea
The positive integers $\{N_i|i=1,\cdots,6\}$ and the  function $f_2(u)$ are given as follows:
\begin{itemize}
\item When $N$ is even, we have
\bea
N_1=N_2=N_5=N_6=\frac32 N, \quad N_3=N_4=2N,
\eea
and the function $f_2(u)$ is
\bea
f_2(u)=1.
\eea

\item When $N$ is odd, we have
\bea N_1=N_2=N_5=N_6=\frac{3N+1}2, \quad
N_3=N_4=2N+1,
\eea
and the functions $f_2(u)$ is
\bea
f_2(u)=\sigma(u).
\eea
\end{itemize}
The associated BAEs (\ref{BAE-1})-(\ref{n-BAE-prod}) become
\bea
\hspace{-0.8truecm}&&e^{2i\pi{l_2}\lambda^{(1)}_j+\phi_2}\frac{Q^{(2)}(\lambda^{(1)}_j+w)}{Q^{(4)}(\lambda^{(1)}_j)}
+c_1e^{2i\pi{l_4}\lambda^{(1)}_j}\frac{a(\lambda^{(1)}_j)}{Q^{(2)}(\lambda^{(1)}_j)}=0,\qquad j=1,\cdots,N_1,\\[6pt]
\hspace{-0.8truecm}&&e^{2i\pi{l_1}\lambda^{(2)}_j+\phi_1}Q^{(1)}(\lambda^{(2)}_j\hspace{-1mm}-\hspace{-1mm}w)\hspace{-1mm}+\hspace{-1mm}
c_1e^{2i\pi{l_4}\lambda^{(2)}_j}\hspace{-1mm}d(\lambda^{(2)}_j)\frac{Q^{(3)}(\lambda^{(2)}_j-w)}{Q^{(1)}(\lambda^{(2)}_j)}=0
,\, j=1,\cdots,N_2,\\[6pt]
\hspace{-0.8truecm}&&e^{2i\pi{l_3}\lambda^{(3)}_j+\phi_{3}}\frac{Q^{(4)}(\lambda^{(3)}_j+w)}{Q^{(6)}(\lambda^{(3)}_j)}
+c_2e^{2i\pi{l_5}\lambda^{(3)}_j}a(\lambda^{(3)}_j)\frac{Q^{(2)}(\lambda^{(3)}_j+w)f_2(\lambda^{(3)}_j)}{Q^{(4)}(\lambda^{(3)}_j)}=0,\no\\[6pt] \hspace{-0.8truecm}&&~~~~~~~~~~~~~~~~~~~~~~~~~j=1,\cdots,N_{3},\\[6pt]
\hspace{-0.8truecm}&&e^{2i\pi{l_2}\lambda^{(4)}_j+\phi_{2}}\frac{Q^{(3)}(\lambda^{(4)}_j-w)}{Q^{(1)}(\lambda^{(4)}_j)}
+c_2e^{2i\pi{l_5}\lambda^{(4)}_j}a(\lambda^{(4)}_j)\frac{Q^{(5)}(\lambda^{(4)}_j-w)f_2(\lambda^{(4)}_j)}{Q^{(3)}(\lambda^{(4)}_j)}=0,\no\\[6pt] \hspace{-0.8truecm}&&~~~~~~~~~~~~~~~~~~~~~~~~~j=1,\cdots,N_{4},\\[6pt]
\hspace{-0.8truecm}&&e^{-2i\pi{\sum^{3}_{k=1}l_k\lambda^{(5)}_j-\sum^{3}_{t=1}(4-t)l_tw-\sum_{l=1}^{3}\phi_l}}Q^{(6)}(\lambda^{(5)}_j+w)\no\\[6pt]
\hspace{-0.8truecm}&&~~~~~~~~~~+c_{3}e^{2i\pi{l_6}\lambda^{(5)}_j}a(\lambda^{(5)}_j)\frac{Q^{(4)}(\lambda^{(5)}_j+w)}{Q^{(6)}(\lambda^{(5)}_j)}=0,
\qquad j=1,\cdots,N_{5},\\[6pt]
\hspace{-0.8truecm}&&e^{2i\pi{l_3}\lambda^{(6)}_j+\phi_{3}}\frac{Q^{(5)}(\lambda^{(6)}_j-w)}{Q^{(3)}(\lambda^{(6)}_j)}
+c_{3}e^{2i\pi{l_{6}}\lambda^{(6)}_j}\frac{a(\lambda^{(6)}_j)}{Q^{(5)}(\lambda^{(6)}_j)}=0,\quad j=1,\cdots,N_{6},
\eea
\bea
\hspace{-0.8truecm}&&\Theta^{(2)}-\Theta^{(1)}=(N_1-N+\frac{N}{4})w+m_1+l_1\tau,\\[6pt]
\hspace{-0.8truecm}&&\Theta^{(1)}-\Theta^{(2)}-\Theta^{(3)}+\Theta^{(4)}=(N_3-N_2+\frac{N}{4})w+m_2+l_2\tau,\\[6pt]
\hspace{-0.8truecm}&&\Theta^{(3)}-\Theta^{(4)}-\Theta^{(5)}+\Theta^{(6)}=(N_5-N_4+\frac{N}{4})w+m_3+l_3\tau,\\[6pt]
\hspace{-0.8truecm}&&-\Theta+\Theta^{(1)}+\Theta^{(2)}-\Theta^{(3)}-N_3w+{3Nw\over4}=m_4+l_4\tau,\\[6pt]
\hspace{-0.8truecm}&&-\Theta-\Theta^{(2)}+\Theta^{(3)}+\Theta^{(4)}-\Theta^{(5)}+N_2w-N_5w+{3Nw\over4}=m_5+l_5\tau,\\[6pt]
\hspace{-0.8truecm}&&-\Theta-\Theta^{(4)}+\Theta^{(5)}+\Theta^{(6)}+N_4w+{3Nw\over4}=m_6+l_6\tau,\\[6pt]
\hspace{-0.8truecm}&&\prod^{N}_{j=1}\frac{Q^{(1)}(\theta_j-w)}{Q^{(2)}(\theta_j)}=e^{-2i\pi{l_1}\Theta-N\phi_1}.
\eea
The main purpose of this Appendix is to show the new features occurred in $Z_4$ case,  which are crucial to understand the structure of the inhomogeneous
$T-Q$ relations (\ref{T-Q-n-1}) for general $Z_n$ case.

\section*{Appendix B: $Z_3$-Belavin model with twisted boundary condition}
\setcounter{equation}{0}
\renewcommand{\theequation}{B.\arabic{equation}}

The Yang-Baxter algebra relation (\ref{Relation1}) and the $Z_n$ symmetry (\ref{Zn-symmetry}) properties of $Z_n$-Belavin R-matrix lead to the fact that the transfer matrix $t^{({\bf \a})}(u)$ given by (\ref{Transfer-twisted}) with different spectral parameters are mutually commuting $[t^{({\bf \a})}(u),t^{({\bf \a})}(v)]=0$. This ensures the integrability of the inhomogeneous $Z_n$-Belavin model with twisted boundary condition.

Without loss of generality, we take the $Z_3$-model with the twisted boundary matrix $G=h$ as an example to construct the solution. The corresponding transfer matrix then reads
\bea
t^{(1,0)}(u)= tr_0\lt(h_0\,T_0(u)\rt).\no
\eea
The invariant relation and operator identities of this transfer matrix $t^{({\bf \a})}(u)$ can be derived in the same way as in dealing with the $su(n)$ spin torus \cite{Hao16}.
The properties of this transfer matrix  imply that the corresponding eigenvalues $\{\Lambda_m(u)|m=1,\cdots,3\}$ satisfy the following functional relations
\bea
\hspace{-1.2truecm}&&\Lambda(\theta_j)\Lambda_m(\theta_j-w)=\Lambda_{m+1}(\theta_j),~~~~~ m=1,2,\quad j=1,\cdots,N,\label{Twisted3-Eigenvalue-1}\\[4pt]
\hspace{-1.2truecm}&&\Lambda_3(u)={\rm Det}_{q}\{h\}{\rm Det}_{q}\{T(u)\}=a(u)d(u-w)d(u-2w),\label{Twisted3-Eigenvalue-function-2}\\[4pt]
\hspace{-1.2truecm}&&\Lambda_2(\theta_j+w)=0,~~~~~~j=1,\cdots,N.\label{Twisted3-Eigenvalue-function-3}
\eea
The periodicity of the $Z_n$-Belavin R-matrix and commuting relation (\ref{gh-relation}) of operators $g$, $h$ give rise to that the eigenvalues are some elliptic polynomials of the fixed degrees $mN$ with the periodicity
\bea
\hspace{-1.2truecm}&&\L_m(u+1)=(-1)^{mN}e^{4mi\pi\over3} \,\L_m(u),\quad m=1,2,\label{Twisted3-Eigenvalue-4}\\[4pt]
\hspace{-1.2truecm}&&\L_m(u+\tau)=(-1)^{mN}\,e^{-2i\pi\{mN(u+\frac{w}{3}+\frac{\tau}{2}-\frac{m-1}{2}w)-m\sum_{l=1}^N\theta_l\}}\,\L_m(u),
\quad m=1,2.\label{Twisted3-Eigenvalue-5}
\eea
Moreover, the unitarity relation (\ref{Unitarity}) and $h^3={\rm id}$ allow us to derive the following identity
\bea
\lt\{\prod_{l=1}^N\,\L(\theta_l)\rt\}^3=\lt\{\prod_{l=1}^N\,a(\theta_l)\rt\}^3.\label{Twisted3-Eigenvalue-6}
\eea

Similar as the periodic case, the relations (\ref{Twisted3-Eigenvalue-1})-(\ref{Twisted3-Eigenvalue-6}) allow us to determine the eigenvalues
$\{\L_m(u)\}$ of the corresponding transfer matrices completely. We can thus express
$\L(u)$ and $\L_2(u)$ in terms of an inhomogeneous $T-Q$ relation as follows. Let us introduce some $Q$-functions
\bea
Q^{(i)}(u)=\prod_{j=1}^{N}\frac{\s(u-\l^{(i)}_j)}{\s(w)},\quad i=1,\cdots,4,\no
\eea parameterized by $4N$ parameters $\{\l^{(i)}_j|j=1,\cdots,N;i=1,\cdots,4\}$ determined later by the associated
BAEs (see below (\ref{Twisted3-BAE-1})-(\ref{Twisted3-BAE-9})). Associated with the
above $Q$-functions, we introduce 5 functions $\{Z_i(u)|i=1,2,3\}$ and $\{X_i(u)|i=1,2\}$ as
\bea
&&Z_1(u)=a(u)e^{2i\pi(l_1+{2\over3})u+\phi_1}\frac{Q^{(1)}(u-w)}{Q^{(2)}(u)},\\[4pt]
&&Z_2(u)=\omega_3\,d(u)e^{2i\pi(l_2+{2\over3}) u+\phi_2}\frac{Q^{(2)}(u+w)Q^{(3)}(u-w)}{Q^{(1)}(u)Q^{(4)}(u)},\\[4pt]
&&Z_3(u)={\omega_3}^2d(u)e^{-2i\pi\{(l_1+l_2+{4\over3})u+(2l_1+l_2+2)w\}-\phi_1-\phi_2}\frac{Q^{(4)}(u+w)}{Q^{(3)}(u)},\\[4pt]
&&X_1(u)=c_1a(u)d(u)e^{2i\pi( l_3+{2\over3})u}\frac{Q^{(3)}(u-w)}{Q^{(1)}(u)Q^{(2)}(u)},\\[4pt]
&&X_2(u)=c_2a(u)d(u)e^{2i\pi(l_4+{2\over3})u}\frac{Q^{(2)}(u+w)}{Q^{(3)}(u)Q^{(4)}(u)},
\eea
where $\omega_3=e^{2i\pi\over3}$, $\{l_i|i=1,\cdots,4\}$ are 4 integers, $\{\phi_i,\,c_i|i=1,2\}$ are 4 complex numbers. Then we can introduce
the  inhomogeneous $T-Q$ relations,
\bea
&&\L(u)=Z_1(u)+Z_2(u)+Z_3(u)+X_1(u)+X_2(u), \label{Twisted3-T-Q-1}\\[4pt]
&&\L_2(u)=Z_1(u)Z_2(u-w)+Z_1(u)Z_3(u-w)+Z_2(u)Z_3(u-w)\no\\[4pt]
&&\qquad\qquad+X_1(u)Z_3(u-w)+Z_1(u)X_2(u-w).\label{Twisted3-T-Q-2}
\eea
In order that the above parameterizations of $\L(u)$ and $\L_2(u)$ become a solution to
(\ref{Twisted3-Eigenvalue-1})-(\ref{Twisted3-Eigenvalue-6}), the $4(N+1)$ parameters $\{\l^{(i)}_j|j=1,\cdots,N; i=1,\cdots,4\}$
and $\{\phi_i,\,c_i|i=1,2\}$ have to satisfy the associated BAEs
\bea
&&\hspace{-1.2truecm}\omega_3 e^{2i\pi (l_2+{2\over3})\l^{(1)}_j+\phi_2}Q^{(2)}(\l^{(1)}_j\hspace{-0.04truecm}+\hspace{-0.04truecm}w)Q^{(2)}(\l^{(1)}_j)
+c_1e^{2i\pi (l_3+{2\over3})\l^{(1)}_j}a(\l^{(1)}_j)Q^{(4)}(\l^{(1)}_j)=0,\no\\
&&\qquad\qquad j=1,\cdots,N,\label{Twisted3-BAE-1}\\[4pt]
&&\hspace{-1.2truecm}e^{2i\pi (l_1+{2\over3})\l^{(2)}_j+\phi_1}Q^{(1)}(\l^{(2)}_j\hspace{-0.04truecm}-\hspace{-0.04truecm}w)Q^{(1)}(\l^{(2)}_j)
\hspace{-0.068truecm}+\hspace{-0.068truecm}c_1e^{2i\pi (l_3+{2\over3})\l^{(2)}_j}d(\l^{(2)}_j)Q^{(3)}(\l^{(2)}_j\hspace{-0.04truecm}-\hspace{-0.04truecm}w)
\hspace{-0.06truecm}=\hspace{-0.06truecm}0,\no\\
&&\qquad\qquad j=1,\cdots,N,\label{Twisted3-BAE-2}\\[4pt]
&&\hspace{-1.2truecm}{\omega_3}^2 e^{-2i\pi\{(l_1+l_2+{4\over3})\l^{(3)}_j+(2l_1+l_2+2)w\}-\phi_1-\phi_2}Q^{(4)}\hspace{-0.06truecm}(\l^{(3)}_j+w)
Q^{(4)}(\l^{(3)}_j)\hspace{-0.068truecm}\no\\
&&\qquad+c_2e^{2i\pi (l_4+{2\over3})\l^{(3)}_j}\hspace{-0.06truecm}a(\l^{(3)}_j)Q^{(2)}(\l^{(3)}_j\hspace{-0.04truecm}+\hspace{-0.04truecm}w)
\hspace{-0.06truecm}=\hspace{-0.06truecm}0, \qquad\qquad j=1,\cdots,N,\label{Twisted3-BAE-3}\\[4pt]
&&\hspace{-1.2truecm}\omega_3 e^{2i\pi (l_2+{2\over3})\l^{(4)}_j+\phi_2}Q^{(3)}(\l^{(4)}_j\hspace{-0.04truecm}-\hspace{-0.04truecm}w)Q^{(3)}(\l^{(4)}_j)
\hspace{-0.068truecm}+\hspace{-0.068truecm}c_2e^{2i\pi (l_4+{2\over3})\l^{(4)}_j}a(\l^{(4)}_j)Q^{(1)}(\l^{(4)}_j)
\hspace{-0.06truecm}=\hspace{-0.06truecm}0,\no\\
&&\qquad\qquad j=1,\cdots,N,\label{Twisted3-BAE-4}
\eea
\bea
&&-\Theta^{(1)}+\Theta^{(2)}-\frac{N}{3}w=m_1+(l_1+{2\over3})\tau,\label{Twisted3-BAE-5}\\[4pt]
&&-\Theta^{(2)}-\Theta^{(3)}+\Theta^{(1)}+\Theta^{(4)}-\frac{N}{3}w=m_2+(l_2+{2\over3})\tau,\label{Twisted3-BAE-6}\\[4pt]
&&-\Theta-\Theta^{(3)}+\Theta^{(1)}+\Theta^{(2)}-\frac{N}{3}w=m_3+(l_3+{2\over3})\tau,\label{Twisted3-BAE-7}\\[4pt]
&&-\Theta-\Theta^{(2)}+\Theta^{(3)}+\Theta^{(4)}+\frac{5N}{3}w=m_4+(l_4+{2\over3})\tau,\label{Twisted3-BAE-8}\\
&&\prod^{N}_{j=1}\frac{Q^{(1)}(\theta_j-w)^3}{Q^{(2)}(\theta_j)^3}=e^{-6i\pi({l_1}+{2\over3})\Theta-3N\phi_1}.\label{Twisted3-BAE-9}
\eea
where $\{m_i|i=1,\cdots,4\}$ are 4 integers and
\bea
\Theta=\sum_{l=1}^N\theta_l,\quad \Theta^{(i)}=\sum_{l=1}^N\l^{(i)}_l,\quad i=1,\cdots,4.\no
\eea

Numerical solutions of the BAEs (\ref{Twisted3-BAE-1})-(\ref{Twisted3-BAE-9}) for small size with random choice of $w$ and $\tau$ imply that 
the Bethe ansatz solution (\ref{Twisted3-T-Q-1}) indeed give the complete solutions of the model. Here we present the numerical solutions of the BAEs for the $N=2$ case in Table  \ref{num-twisted-N2};
The eigenvalue calculated from (\ref{Twisted3-T-Q-1}) is the same as that from the exact diagonalization of the Hamiltonian (\ref{Ham})  with the twisted boundary condition  (\ref{Twisted-condition})  associated with $G=h$.
\begin{landscape}

\begin{table}

\caption{Solutions of BAEs (\ref{Twisted3-BAE-1})-(\ref{Twisted3-BAE-9}) for the $Z_3$-Belavin model with the twisted boundary condition, $N=2$, $\{\theta_j\}=0$, $w=-0.5$, $\tau=i$ and the parameters $m_1=1$, $m_2=m_3=m_4=l_1=l_2=l_3=l_4=0$. The symbol $m$ indicates the number of the  eigenenergy $E$. }\label{num-twisted-N2}
\begin{tabular}{|c|c|c|c|c|}
\hline\hline $\lambda^{(1)}_1$ & $\lambda^{(1)}_2$ & $\lambda^{(2)}_1$ & $\lambda^{(2)}_2$ & $\lambda^{(3)}_1$ \\ \hline
$-0.2987+0.0905i$  &  $0.2987-0.0905i$  &  $0.8147+0.0484i$  &  $-0.1481+0.6183i$  &  $0.3220+0.1800i$ \\
$0.7013+0.0905i$  &  $-0.7013-0.0905i$  &  $1.8147+0.0484i$  &  $-1.1481+0.6183i$  &  $-0.3220-0.1800i$ \\
$0.3498+0.1256i$  &  $-0.3498-0.1256i$  &  $-0.1853+1.0484i$  &  $0.8519-0.3817i$  &  $0.3436-0.1110i$ \\
$-0.3220-0.1800i$  &  $0.3220+0.1800i$  &  $0.1667+0.6667i$  &  $0.5000-0.0000i$  &  $-0.7013-0.0905i$ \\
$0.6780-0.1800i$  &  $-0.6780+0.1800i$  &  $1.1667-0.3333i$  &  $-0.5000+1.0000i$  &  $-0.7013-0.0905i$ \\
$0.6780-0.1800i$  &  $-0.6780+0.1800i$  &  $1.1667-0.3333i$  &  $-0.5000+1.0000i$  &  $-0.7013-0.0905i$ \\
$1.2648+0.0940i$  &  $-1.2648-0.0940i$  &  $-0.5215-0.3974i$  &  $1.1882+1.0641i$  &  $1.7274+0.1920i$ \\
$-0.3436+0.1110i$  &  $0.3436-0.1110i$  &  $-1.8118+0.0641i$  &  $2.4785+0.6026i$  &  $1.3498+0.1256i$ \\
$2.6564+0.1110i$  &  $-2.6564-0.1110i$  &  $-0.8118+0.0641i$  &  $1.4785+0.6026i$  &  $-0.3498-0.1256i$ \\
\hline\hline
$\lambda^{(3)}_2$ & $\lambda^{(4)}_1$ & $\lambda^{(4)}_2$ & $\phi_1$ & $\phi_2$ \\ \hline
$0.6780-0.1800i$  &  $1.0000+0.0000i$  &  $0.3333+1.3333i$  &  $1.4622-1.0718i$  &  $4.5236-3.7306i$ \\
$1.3220+0.1800i$  &  $1.0000+1.0000i$  &  $0.3333+0.3333i$  &  $1.4622+9.4002i$  &  $2.4292-16.2969i$ \\
$0.6564+0.1110i$  &  $0.0215+1.3974i$  &  $1.3118-0.0641i$  &  $4.3225+9.5401i$  &  $2.5056-12.6652i$ \\
$1.7013+0.0905i$  &  $0.6853+0.9516i$  &  $0.6481+0.3817i$  &  $1.7970-6.8967i$  &  $1.7596+3.7306i$ \\
$1.7013+0.0905i$  &  $0.6853+0.9516i$  &  $0.6481+0.3817i$  &  $3.8914+18.2360i$  &  $-0.3348-10.9302i$ \\
$1.7013+0.0905i$  &  $0.6853+0.9516i$  &  $0.6481+0.3817i$  &  $3.8914+14.0472i$  &  $-0.3348-15.1190i$ \\
$-0.7274-0.1920i$  &  $-1.0000+1.0000i$  &  $2.3333+0.3333i$  &  $4.2878+4.8973i$  &  $-0.6765+4.5508i$ \\
$-0.3498-0.1256i$  &  $1.6481+0.3817i$  &  $-0.3147+0.9516i$  &  $1.8336-2.8543i$  &  $1.8807-8.7834i$ \\
$1.3498+0.1256i$  &  $2.6481+0.3817i$  &  $-1.3147+0.9516i$  &  $1.8336+18.0896i$  &  $1.8807-19.2553i$ \\
\hline\hline
$c_1$ & $c_2$ & $E$ & $m$ & \\ \hline
$3.4493-2.9903i$  &  $34.0526+77.9548i$ & $-5.0705-0.1943i$ & $1$ & \\
$-4.3143-1.4921i$  &  $4.1934+9.5997i$ & $-5.0705-0.1943i$ & $1$ & \\
$-102.6655-26.2513i$  &  $2.4711-8.2348i$ & $-5.0705-0.1943i$ & $1$ & \\
$3.1926-4.5605i$  &  $-6.1152+0.6177i$ & $0.8364+1.3278i$ & $2$ & \\
$25.9258-37.0339i$  &  $0.4424+0.6141i$ & $0.8364+1.3278i$ & $2$ & \\
$19.1094+40.9693i$  &  $-0.7531+0.0761i$ & $0.8364+1.3278i$ & $2$ & \\
$24.2590-49.1261i$  &  $-0.6505-0.0544i$ & $4.2341-1.1336i$ & $3$ & \\
$-4.3389-0.6734i$  &  $-8.8842+2.4665i$ & $4.2341-1.1336i$ & $3$ & \\
$2.7526-3.4208i$  &  $2.3061-8.9272i$ & $4.2341-1.1336i$ & $3$ & \\
\hline\end{tabular}

\end{table}
\end{landscape}

\end{document}